\documentclass[aps,prl,reprint,groupedaddress,showpacs,amsmath,amssymb,twocolumn,floatfix]{revtex4-1}
\usepackage{graphicx}
\usepackage{bm}
\usepackage{color}
\usepackage[normalem]{ulem}
\usepackage{amsmath}
\usepackage{tabularx}

\usepackage[colorlinks=true]{hyperref}  
\hypersetup{
    unicode=false,          
    pdftoolbar=true,        
    pdfmenubar=true,        
    pdffitwindow=false,     
    pdfstartview={FitH},    
    pdftitle={Angle-resolved nonreciprocity},    
    pdfauthor={},     
    pdfsubject={},   
    pdfcreator={},   
    pdfproducer={}, 
    pdfkeywords={} {} {}, 
    pdfnewwindow=true,      
    colorlinks=true,       
    linkcolor=magenta, 
    citecolor=blue,        
    filecolor=magenta,      
    urlcolor=blue           
} 
\usepackage{cleveref}

\newcommand{\tlg}{tTLG}
\newcommand{\WSe}{WSe$_2$}

\newcommand{\Vto}{$V^{2\omega}_{\perp}$}
\newcommand{\Vpo}{$V^{2\omega}_{\parallel}$}

\usepackage{bbm}

\renewcommand{\vec}[1]{\boldsymbol{#1}}

\begin{document}

\title{Angle-resolved transport nonreciprocity and spontaneous symmetry breaking in twisted trilayer graphene}

\author{Naiyuan James Zhang$^{1}$}
\author{Jiang-Xiazi Lin$^{1}$}
\author{Dmitry V. Chichinadze$^{2}$}
\author{Yibang Wang$^{1}$}
\author{Kenji Watanabe$^{3}$}
\author{Takashi Taniguchi$^{4}$}
\author{Liang Fu$^{5}$}
\author{J.I.A. Li$^{1}$}
\email{jia\_li@brown.edu}

\affiliation{$^{1}$Department of Physics, Brown University, Providence, RI 02912, USA}
\affiliation{$^{2}$ National High Magnetic Field Laboratory, Tallahassee, Florida, 32310, USA}
\affiliation{$^{3}$Research Center for Functional Materials, National Institute for Materials Science, 1-1 Namiki, Tsukuba 305-0044, Japan}
\affiliation{$^{4}$International Center for Materials Nanoarchitectonics,
National Institute for Materials Science,  1-1 Namiki, Tsukuba 305-0044, Japan}
\affiliation{$^{5}$Department of Physics, Massachusetts Institute of Technology, Cambridge, MA 02139, USA}

\date{\today}

\maketitle

\textbf{The ability to identify and characterize spontaneous symmetry breaking is central to our understanding of 2D materials with strong correlation, such as the moir\'e flat bands in magic-angle twisted graphene bilayer and trilayer.  In this work, we utilize angle-resolved measurements of transport nonreciprocity to investigate spontaneous symmetry breaking in twisted trilayer graphene. By analyzing the angular dependence of nonreciprocity in both longitudinal and transverse channels, we are able to identify the symmetry axis associated with the underlying electronic order. We report that a hysteretic rotation in the mirror axis can be induced by thermal cycles and a large current bias, which offers unambiguous evidence for the spontaneous breaking of rotational symmetry.  Moreover, the onset of nonreciprocity with decreasing temperature coincides with the emergence of orbital ferromagnetism. Combined with the angular dependence of the superconducting diode effect, our findings uncover a direct link between rotational and time-reversal symmetry breaking. These symmetry requirements point towards the exchange-driven instabilities in the momentum space as a possible origin for transport nonreciprocity in tTLG.
}

\begin{figure*}[!t]
\includegraphics[width=0.63\linewidth]{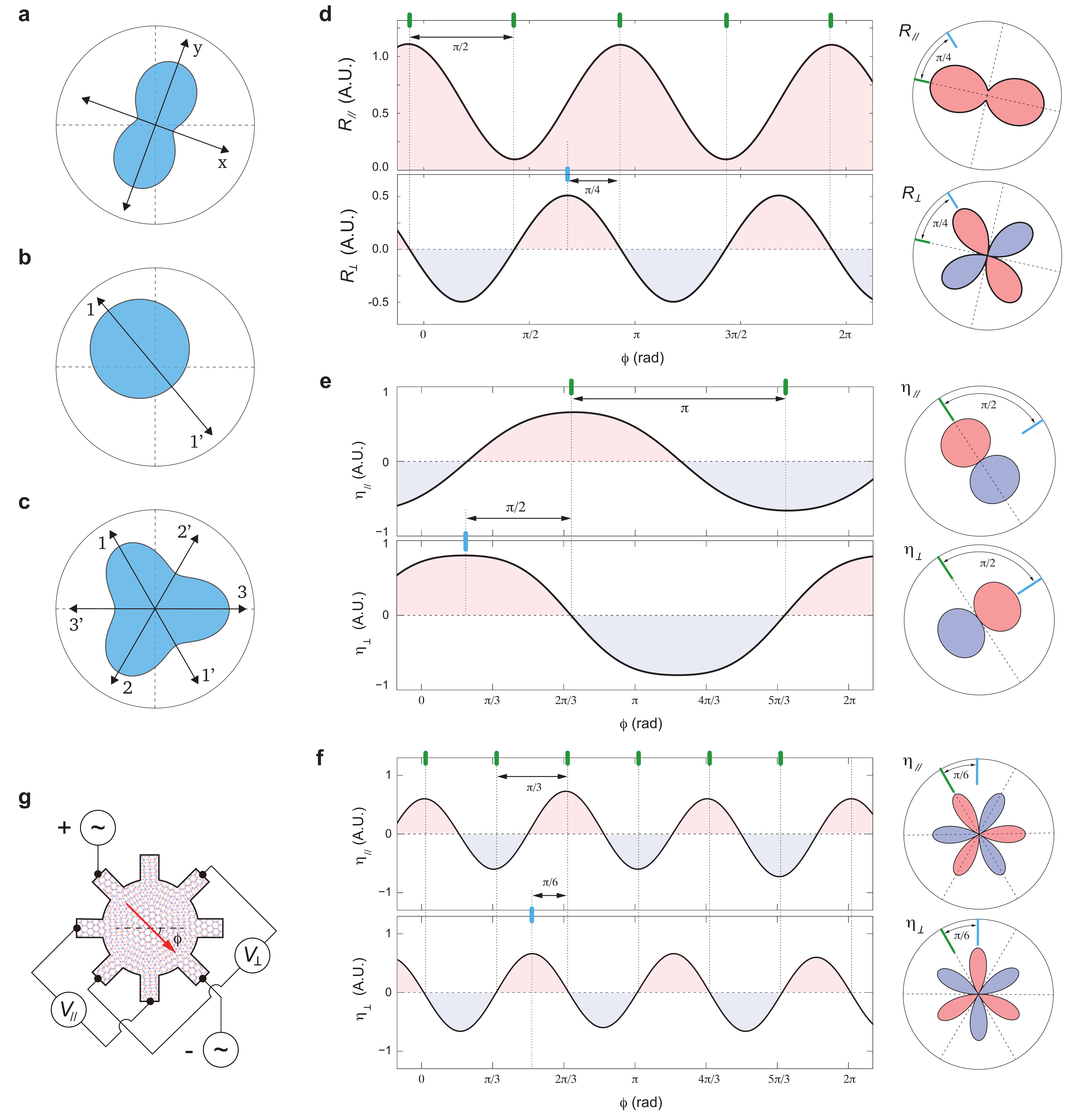}
\caption{\label{fig1} {\bf{Symmetry analysis of the angle-resolved transport response.}} (a-c)  Schematic diagram of Fermi surface contour with different angular symmetries. Black arrows mark the azimuth direction of the mirror axis. (d-f) Expected angular dependence in the transport response when the underlying electronic state is described by (d) one, (e) two, and (f) three mirror axes. Left (right) panels show the same angular dependence in the Cartesian (Polar) coordinate. Panel (d) plots the angular dependence of longitudinal and transverse resistance, $R_{\parallel}$ and $R_{\perp}$, which are measured with a small DC current bias. panel (e) and (f) represents the angular dependence of transport nonreciprocity. 
 (g) Schematic diagram of the ``sunflower'' shaped sample and the ARNTM setup.
} 
\end{figure*}

\begin{figure*}
\includegraphics[width=0.95\linewidth]{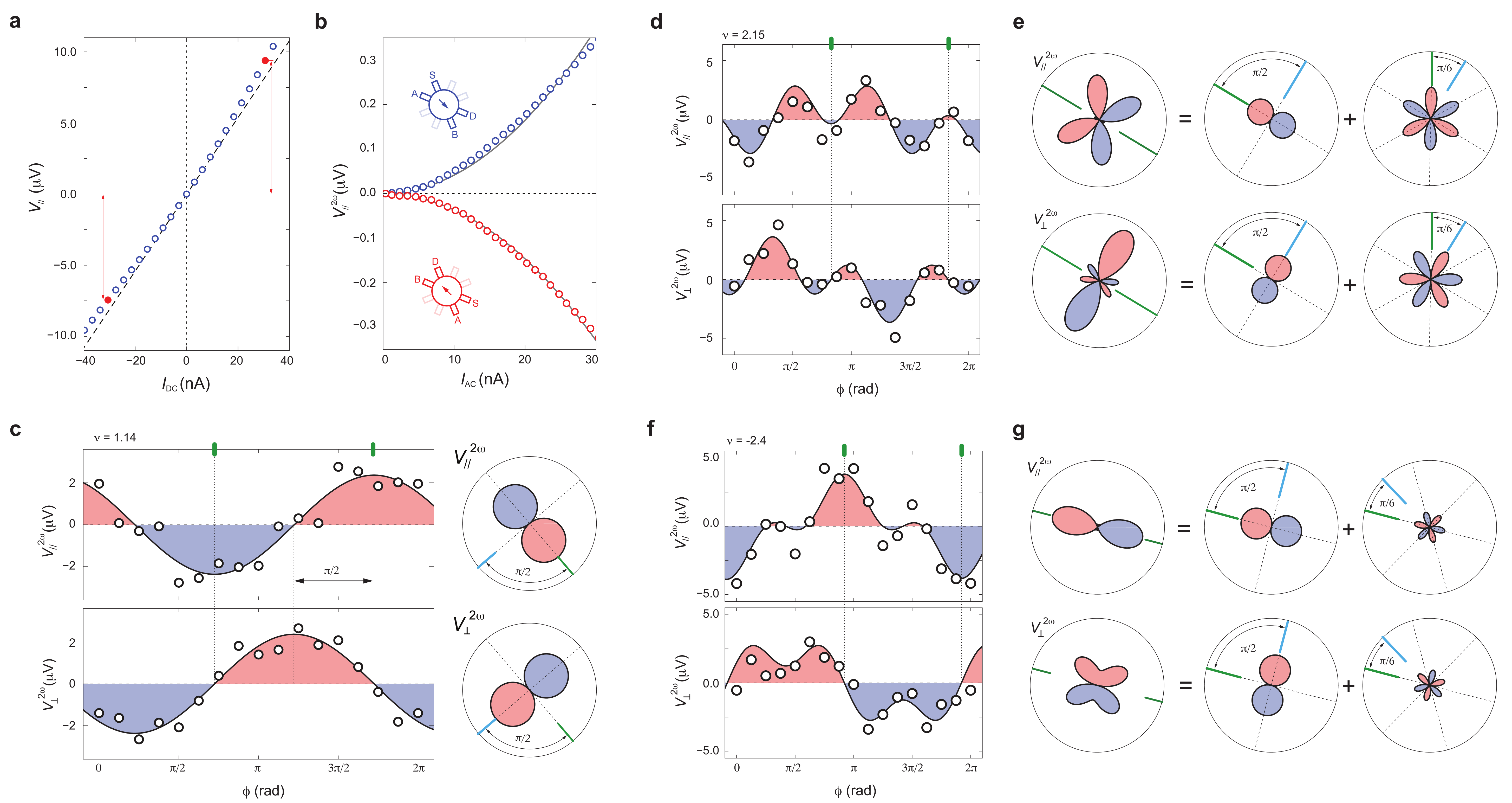}
\caption{\label{fig2} {\bf{Transport nonreciprocity in tTLG.}} (a) Current-voltage characteristic measured with a DC current bias. The black dashed line represents the linear component of the transport response. (b) The second-harmonic nonlinear transport response \Vpo\ as a function of $I_{\textrm{AC}}$, measured with current flowing along $\phi = 135^{\circ}$ (blue) and $315^{\circ}$ (red). (c) The angular dependence of \Vpo\ and \Vto\ measured at $\nu=1.14$.  (d-e) The angular dependence of \Vpo\ and \Vto\ at $\nu=2.15$ in the (d) Cartesian and (e) Polar axis. (f-g) The angular dependence of \Vpo\ and \Vto\ at $\nu=-2.4$  in the (f) Cartesian and (g) Polar axis. The left panel of  (e) and (g) displays the best angular fit based on Eq.~1, which is a linear combination of a one-fold (middle panel) and a three-fold components (right panel). Green solid line marks the direction of the mirror axis.
Measurements in (c-g) are performed at $I_{AC} =100$ nA. All measurements are performed at $T=20$ mK and $B=0$. 
}
\end{figure*}

\begin{figure*}
\includegraphics[width=0.9\linewidth]{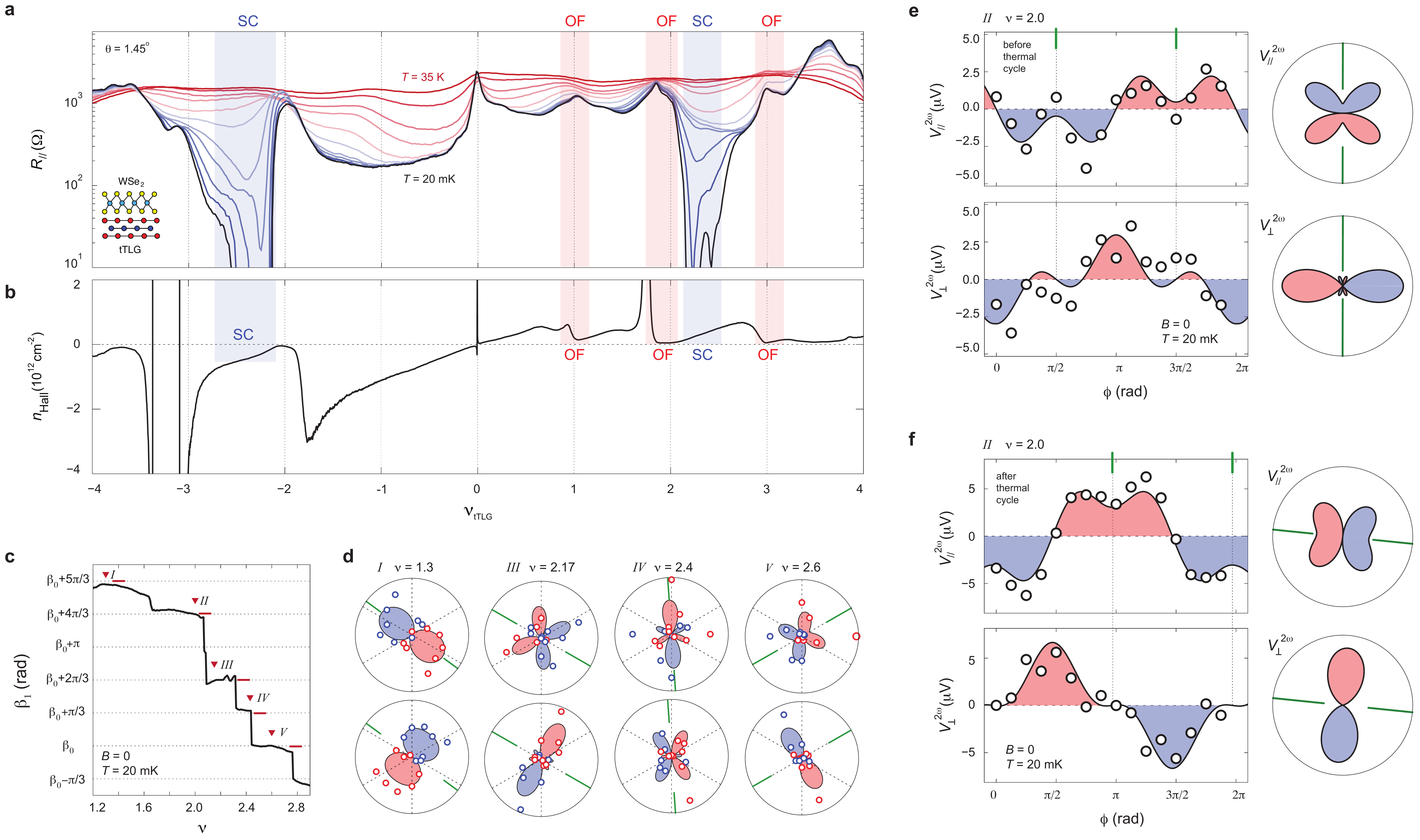}
\caption{\label{fig3}  {\bf{Moir\'e flat band and the density dependence of nonreciprocity.}} (a) Longitudinal resistance $R_{\parallel}$ and (b) Hall density $n_{Hall}$ as a function of moir\'e band filling $\nu$, measured at $T=20$ mK and $I_{DC}=5$ nA. The density regimes of the superconducting phase (SC) and the orbital ferromagnetic (OF) order are marked with blue and red vertical stripes, respectively.  Inset in (a) shows the schematic diagram of sample composition, where a tTLG is proximitized with a few layer tungsten diselenide (\WSe) crystal.
(c) The mirror axis orientation $\beta$ as a function of moir\'e band filling. (d) Polar-coordinate plots of the angular dependence of transport nonreciprocity measured from plateau \textit{I}, \textit{III}, \textit{IV}, and \textit{V} in (a). (e) and (f) The angular dependence of\ \Vpo\ and \Vto\ measured at $\nu=2.0$, which is on plateau \textit{II}, (e) before and (f) after a thermal cycle. Measurements in (c-f) are performed at $T=20$ mK, $B=0$, and $I_{AC}=100$ nA.
}
\end{figure*}

Electronic nematicity describes the Coulomb-driven phenomenon with reduced in-plane rotational symmetry compared to the underlying crystal lattice ~\cite{Fradkin2010nematic,Oganesyan2001nematic,Kivelson1998nematic}. For instance, the orthorhombic anisotropy emerges as the Fermi surface undergoes a quadruple distortion to save Coulomb energy. Such a Coulomb-driven phenomenon has been widely observed in a series of strongly correlated 2D systems, such as strontium ruthenate, cuprate materials ~\cite{Wu2017nematic,Wu2020nematic,Ando2002nematic,Hinkov2008nematic}, magic-angle twisted bilayer and trilayer graphene ~\cite{Jiang2019STM,Choi2019STM,Kerelsky2019STM,Cao2020nematicity,Rubio2022nematic,Zhang2022sunflower}. An orthorhombic anisotropy features two mirror axes that are orthogonal to each other (Fig.~\ref{fig1}a).  As a result, the symmetry associated with a two-fold in-plane rotation, $C_2$, is preserved. Further lifting $C_2$ symmetry promises to unlock a new dimension of rich physical constructions beyond the orthorhombic anisotropy. It is recently proposed that exchange interaction amongst trigonally-warped Fermi pockets enables novel instabilities in the momentum-space. For instance, electronic state in multilayer graphene could acquire  a non-zero net momentum as charge carriers spontaneously condense into one of the Fermi pockets ~\cite{Lin2023BLG,Dong2021momentum,Jung2015momentum,Huang2022momentum}. Unlike the orthorhombic anisotropy, the momentum-space instability breaks $C_2$ and time-reversal symmetry $T$ simultaneously. 

Rotational symmetry breaking can be identified based on angle-resolved transport measurement ~\cite{Wu2017nematic,Wu2020nematic}, which conventionally focuses on the linear and reciprocal component of transport response, such as the longitudinal and transverse resistance, $R_{\parallel}$ and $R_{\perp}$, in the limit of small current bias. 
The angular dependence of $R_{\parallel}$ and $R_{\perp}$ is constrained by the underlying symmetry axes.  A Fermi surface with quadruple distortion preserves $C_2$ symmetry, which gives rise to two orthogonal mirror axes, which are marked by black arrows in Fig.~\ref{fig1}a. As a function of $\phi$, the longitudinal (transverse) transport response is expected to be symmetric (anti-symmetric) around a mirror axis. Not only does this symmetry constraint gives rise to a period of $\pi$ in the angular oscillation of $R_{\parallel}$ and $R_{\perp}$, but it also accounts for a phase shift of $\pi/4$ between longitudinal and transverse channels. Together, the angular dependence can be expressed as $R_{\parallel} = A \cos(2\phi - \alpha)$ and $R_{\perp} = A \sin(2\phi - \alpha)$. Here, $\alpha$ defines the orientation of the mirror axes, whereas the phase shift of $\pi/4$ is captured by cosine and sine functions.

Beyond the orthorhombic anisotropy, the nature of electronic orders with $C_2$ symmetry breaking is directly linked to the nonlinear and nonreciprocal transport response. For simplicity, we label nonreciprocity in the longitudinal (transverse) channels as $\eta_{\parallel}$ ($\eta_{\perp}$),  which is simply defined as the difference in $R_{\parallel}$ ($R_{\perp}$) between forward and reverse current bias. The angular dependence of nonreciprocity shares the same symmetry constraint as the linear transport response. Consequently, $\eta_{\parallel}$ ($\eta_{\perp}$) is expected to be maximized (zero) when current flows along the mirror line. Owing to $C_2$ symmetry breaking, the electronic order can have either one or three mirror axes, as shown in Fig.~\ref{fig1}b-c). 
An electronic state with a single mirror axis (Fig.~\ref{fig1}b) enables an angular period of $2\pi$ and a phase shift of $\pi/2$ between $\eta_{\parallel}$ and $\eta_{\perp}$ (Fig.~\ref{fig1}e). Along the same vein, a state with three mirror axes (Fig.~\ref{fig1}c) is identifiable by a period of $2\pi/3$ in the angular oscillation, along with a phase shift of $\pi/6$ between $\eta_{\parallel}$ and $\eta_{\perp}$ (Fig.~\ref{fig1}f).

In this work, we investigate the angular dependence of transport nonreciprocity in mirror-symmetric twisted trilayer graphene (tTLG), which sheds new light on the nature of spontaneous symmetry breaking in a moir\'e flat band. 
To perform the angle-resolved transport measurement, we shape tTLG samples into the ``sunflower'' geometry, as shown in Fig.~\ref{fig1}g. By using different contact pairs as source and drain, the ``sunflower'' geometry allows us to flow current is $16$ azimuth directions in the range of $0 < \phi < 2\pi$ (see Fig.~\ref{figSetup} for different measurement configurations). This provides the necessary resolution to identify angular oscillations with a period of $2\pi$ and $2\pi/3$. When current flows along the azimuth direction $\phi$, voltage responses $V_{\parallel}(\phi)$ and $V_{\perp}(\phi)$ are measured between contact leads that are aligned parallel and perpendicular to the current flow direction, as shown in Fig.~\ref{fig1}g. Longitudinal and transverse resistance are extracted as $R_{\parallel}(\phi) = V_{\parallel}(\phi)/I_{DC}$ and $R_{\perp}(\phi) = V_{\perp}(\phi)/I_{DC}$, whereas nonreciprocity is defined as $\eta_{\parallel} (\phi)= V_{\parallel}(\phi) - V_{\parallel}(\phi+\pi)$, $\eta_{\perp} (\phi)= V_{\perp}(\phi) - V_{\perp}(\phi+\pi)$. 

The presence of transport nonreciprocity is demonstrated by the current-voltage (I-V) characteristics in Fig.~\ref{fig2}a, where the voltage response at large current deviates from the linear component (the black dashed line). The nonreciprocity, extracted by subtracting the linear component, exhibits a quadratic current dependence (see Fig.~\ref{DCEta}). In the presence of an AC current bias with frequency $\omega$, the nonreciprocal response is proportional to $sin^2(\omega t) \sim sin (2\omega t)$. As such, $\eta$ can be conveniently probed as the amplitude of the AC response at the second-harmonic frequency $2\omega$. Fig.~\ref{fig2}b plots the second-harmonic voltage response, $V_{\parallel}^{2\omega}$, which exhibits the quadratic dependence on the AC current bias. A rotation of $180^{\circ}$ in the measurement configuration induces a sign reversal in $V_{\parallel}^{2\omega}$. Since the sample geometry respects the rotation, the sign reversal offers a strong indication of two-fold rotational symmetry breaking. We note that nonreciprocity extracted from the DC I-V curve is equivalent to the second-harmonic nonlinear response (Fig.~\ref{ACDCcompare}). 

Fig.~\ref{fig2}c-g plots the angular dependence of nonreciprocity measured at different moir\'e band fillings. At $\nu = 1.14 $, both \Vpo\ and \Vto\ are captured by cosine functions with a period of $2\pi$. Fig.~\ref{fig2}c also displays a relative phase shift of $\pi/2$ across two channels. This is consistent with the presence of one mirror axis, which is marked by the green solid line in Fig.~\ref{fig2}c. In comparison, the angular dependence in Fig.~\ref{fig2}d and f exhibits a more complex functional form, which reveals a unique connection between longitudinal and transverse channels. Fig.~\ref{fig2}e and g plots the polar-coordinate plot of the best angular fit for Fig.~\ref{fig2}d and f (left panel), which is a linear combination of the one-fold (middle panel) and three-fold (right panel) components. Green and blue solid lines mark the phases of the longitudinal and transverse channels. Between these two channels, the one-fold components display a phase shift of $\pi/2$   whereas the phases of three-fold components are offset by $\pi/6$.

The unique connection between longitudinal and transverse channels provides a strong constraint for identifying the functional form for the angular dependence of nonreciprocity. Both channels of nonreciprocity can be simultaneously captured by a rather simple expression,
\begin{eqnarray}
    V_{\parallel}^{2\omega}(\phi) &=& V_1 \cos(\phi-\beta)+V_3 \cos(3(\phi-\beta)), \nonumber\\ 
    V_{\perp}^{2\omega}(\phi) &=& V_1 \sin(\phi-\beta)+V_3 \sin(3(\phi-\beta)).
\end{eqnarray}
\noindent 
Here $\beta$ defines the azimuth orientation of the mirror axis. $V_1$ ($V_3$) denotes the amplitude of the one-fold (three-fold) component of the angular oscillation, whereas the phase shift between longitudinal and transverse channels is naturally captured by cosine and sine. 
Eq.~1 resembles a Fourier expansion around a mirror axis, where the longitudinal (transverse) nonreciprocity is expressed by the linear combination of $\cos (N\phi)$ ($\sin (N\phi)$). While an expansion allows any odd integer values of $N$, only $N=1$ and $3$ are observed in our measurement. Fig.~\ref{fig10petal} shows nonreciprocity measured with a higher angular resolution, which confirms that the dominating role of the $N=1$ and $3$ components. 

The quality of the angular fit can be gauged based on the relative root mean square error (RRMSE), as shown in Fig.~\ref{figError} and Fig.~\ref{figError3fold}. RRMSE exhibits sharp minima around the optimal fitting parameter, suggesting that $V_1$, $V_3$ and $\beta$ are uniquely determined. RRMSE of the best fit is around $0.05$ for the angular dependence in Fig.~\ref{fig2}. This is a strong indication that Eq.~1 offers a good description for the angular dependence of nonreciprocity.  
The ``sunflower'' geometry also enables measurement and analysis schemes beyond the simple angle-resolved transport. These schemes offer direct assessment for the uniformity of the sample. Theoretically,  the potential distribution across the circumference of the sample is solely determined by the conductivity matrix ~\cite{Vafek2022sunflower}. This distribution can be mapped through the linear transport response from an extensive series of measurement configurations, as shown in Fig.~\ref{Fullfit} ~\cite{Zhang2022sunflower}. By fitting the potential distribution, a single conductivity matrix is extracted with a RRMSE of $0.017$. The excellent fit quality points towards a uniform transport response across the entire sample, which indicates that the impact from moir\'e disorder and inhomogeneity is negligible.

Next, we will examine the linear  transport response of tTLG. As shown in Fig.~\ref{fig3}a-b, the tTLG sample exhibits a series of hallmark signature of a moir\'e flat band. For instance, resistance peaks in $R_{\parallel}$ is observed near $\nu_{tTLG} = -2$, $+1$, $+2$, and $+3$ (Fig.~\ref{fig3}a), which coincides with resets in the Hall density $n_{Hall}$ (Fig.~\ref{fig3}b). This is consistent with previous observations from magic-angle twisted graphene bilayer and trilayer, where valley and spin degeneracy is spontaneously lifted near integer band filling, giving rise to a cascade of Fermi surface reconstructions ~\cite{Zondiner2020cascade,Park2021flavour,Xie2021cascade,Kang2021cascades,Liu2022DtTLG}. 
As shown in the inset of Fig.~\ref{fig3}a, the tTLG is proximitized with a \WSe\ crystal. Fig.~\ref{figComparison} investigates the influence of the proximity effect through comparison with another tTLG sample without the \WSe\ crystal. In both samples, Fermi surface reconstruction, evidenced by resistance peak and Hall density reset, occurs at the same band fillings. Superconductivity in both samples emerges near half-filling of electron- and hole-doped bands. Moreover, the band-filling-temperature ($\nu-T$) maps from both samples reveal the same resistance oscillation at high temperature. Combined, these observations suggest that the nature of isospin degeneracy lifting is unaffected by the proximity effect. The orbital ferromagnetic order, however, is only observed in the tTLG sample with the proximity effect. This is consistent with previous observations of proximity-induced anomalous Hall effect in magic-angle twisted bilayer graphene ~\cite{Lin2022SOC}.

Beyond the sequence of Fermi surface reconstructions, angle-resolved nonreciprocity measurement reveals a new type of cascade phenomenon. Fig.~\ref{fig3}c and d shows the angular dependence of \Vpo\ and \Vto\ in the density regime of $1.2 < \nu_{tTLG} < 2.8$. The orientation of the mirror axis, marked by the green solid line, is determined by  the best angular fit based on Eq.~1. As a function of band filling, $\beta$ exhibits a series of plateaus  (black solid line in Fig.~\ref{fig3}c), which point towards a cascade of rotations in the mirror axis. 
The values of $\beta$ between adjacent plateaus, marked as \textit{I} through \textit{V}, are offset by integer multiple of $\pi/3$. This defines six  characteristic directions, highlighting a possible link to the crystallographic axes of the underlying moir\'e lattice. 

According to the angular dependence of nonreciprocity in Fig.~\ref{fig3}d, the underlying electronic order features a single mirror axis, which is indicative of broken three-fold rotational symmetry $C_3$. That the mirror axis rotates with varying moir\'e band filling stipulates that the $C_3$ breaking cannot be accounted for by the influence of hetero-strain alone. In addition, we show that a hysteretic rotation in the mirror axis can be induced by thermal cycling the sample, or the application of a large current bias. Fig.~\ref{fig3}e-f compares the angular dependence of nonreciprocity measured at the same moir\'e filling from two consecutive cool down cycles. The best angular fit with Eq.~1 reveals a prominent rotation in the mirror axis before (Fig.~\ref{fig3}e) and after the thermal cycle (Fig.~\ref{fig3}f). Similarly, the application of a large current bias also enables a prominent rotation in the mirror axis, as shown in (Fig.~\ref{CurrentHysteresis}b. Such hysteretic transition is associated with a hysteresis loop in the second-harmonic nonlinear response as current bias is swept back and forth (Fig.~\ref{CurrentHysteresis}c). These hysteretic transitions point towards nearly degenerate electronic orders with different mirror axis orientations. The degeneracy is lifted by the process of spontaneous rotational symmetry breaking.

\begin{figure*}
\includegraphics[width=0.88\linewidth]{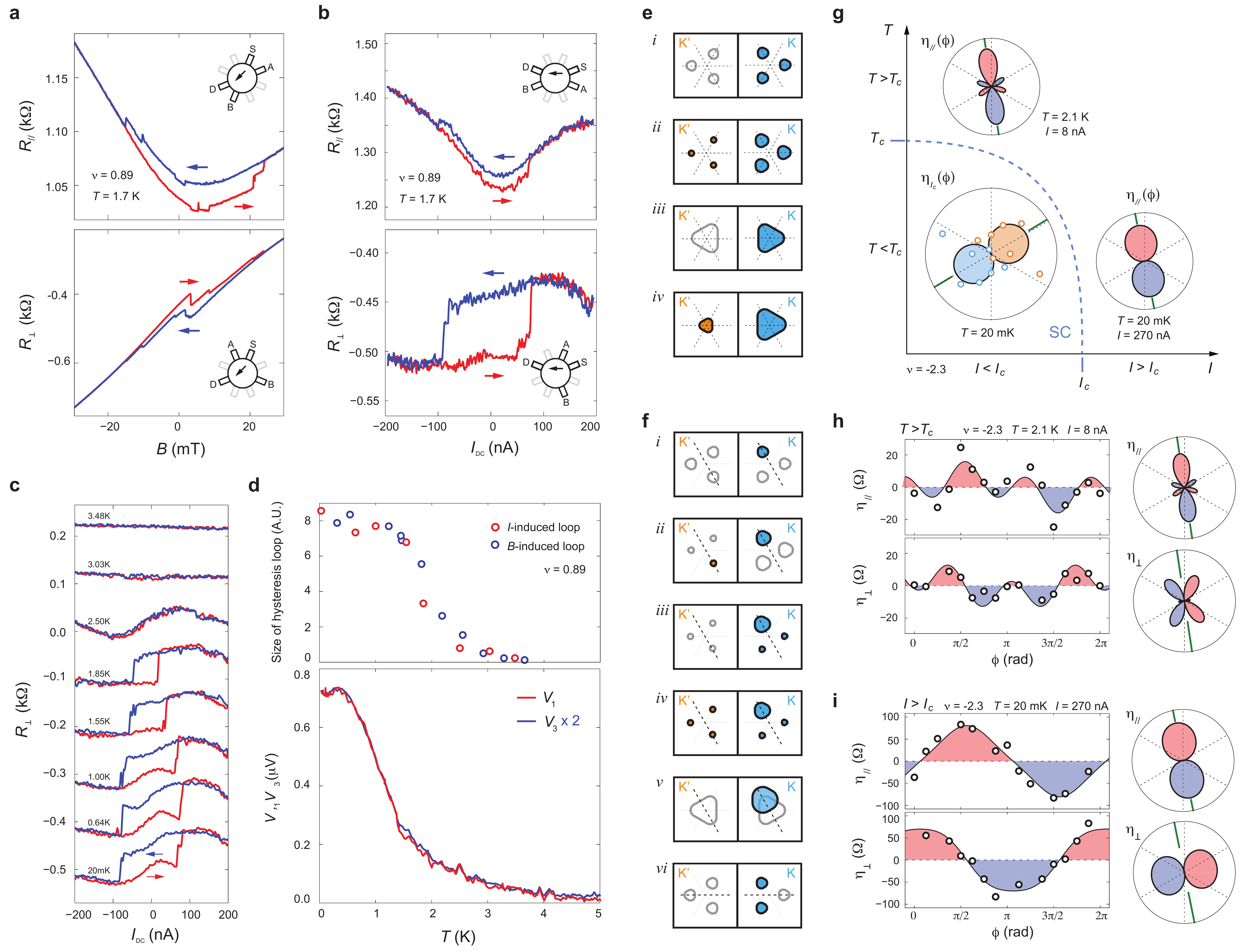}
\caption{\label{fig4} {\bf{Nonreciprocity, orbital ferromagnetism and superconductivity.} } 
(a-b) Hysteretic transitions at $\nu=0.89$ is induced by sweeping (a) magnetic field $B$ and  (b) DC current bias $I_{DC}$ back and forth. Top and bottom panels show longitudinal and transverse resistance, $R_{\parallel}$ and  $R_{\perp}$, respectively.  
(c) Current-induced hysteresis loops measured at different temperatures. (d) Size of the hysteresis loops (top panel) is compared against the temperature dependence of $V_1$ and $V_3$  (bottom panel), which are extracted from the angular oscillation in \Vpo\ and \Vto\ according to Eq.~1. 
(e-f) Schematic diagram of Fermi surface occupation across valley $K$ and $K'$ in the presence of (e)  valley polarization with three mirror axes and (f) momentum polarization with one mirror axis. The spin degree of freedom is omitted. (g) Schematic current-temperature (I-T) phase diagram at the optimal doping of the  superconducting phase. The blue dashed line marks the SC-normal state transition. The angular dependence of nonreciprocity is shown for the superconducting ($\eta_{I}(\phi)$) and normal phases ($\eta_{\parallel}(\phi)$). See Fig.~\ref{EtaIc}) for more details on the nonreciprocity in the superconducting transport.
(h-i) The angular dependence of nonreciprocity, $\eta_{\parallel}$ (top panels) and $\eta_{\perp}$ (bottom panels), measured in the normal state at (g) $T > T_c$, and (h) $I > I_c$. Measurements are performed at the optimal density of the superconducting phase $\nu=-2.3$. 
} 
\end{figure*}

The foremost question is the microscopic origin underlying the rotational symmetry breaking. To address this question, we first consider the requirements of $C_2$ symmetry breaking in association with time-reversal symmetry $T$. There are two possible scenarios (see detailed discussion in SI): (i) $C_2$ is broken and $T$ is preserved. This scenario can be enabled by Berry curvature dipole and skew scattering ~\cite{Ma2019nonlinear,Kang2019nonlinear,He2022nonlinear,Sinha2022nonlinear};
(ii) $C_2$ and $T$ breaking occurs simultaneously. Owing to the influence of trigonal warping in tTLG ~\cite{Khalaf2019,Hao2021tTLG}, $C_2$ and $T$ breaking are naturally realized by the presence of valley polarization, which describes an imbalanced charge carrier occupation of Fermi surface across different corners of the Brillouin zone  ~\cite{Novoselov2004,novoselov2005two,Geim2013RV,Novoselov2016RV,Kim2016RV}. In the presence of sublattice polarization, valley polarization is directly linked to an orbital ferromagnetic order, which is identifiable through the hysteretic transitions of the anomalous Hall effect ~\cite{Sharpe2019,Serlin2019,Lin2022SOC}. 

Given the hysteretic rotation in the mirror axis and the evolution with moir\'e band filling, the angular dependence of nonreciprocity likely cannot be accounted for by Berry curvature dipole or skew scattering. Therefore,  we examine the potential role of valley polarization and orbital ferromagnetism. In the tTLG sample, orbital ferromagnetic order is observed near integer fillings of the electron-doped band.
Fig.~\ref{fig4}a-b plots transport response measured at $\nu = 0.89$ as a function of magnetic field $B$ and DC current bias $I_{DC}$. As $B$ and $I_{DC}$ are swept back and forth, hysteretic transitions are observed in both $R_{\parallel}$ (top panels) and $R_{\perp}$ (bottom panels). This is in excellent agreement with the anomalous Hall effect observed in other multilayer graphene heterostructures~\cite{Serlin2019,Chen20201N2,Polshyn20201N2,Lin2022SOC}. Owing to the presence of a Dirac-like band, tTLG sample is always metallic, even when the moir\'e flat band develops an energy gap with the emergence of orbital ferromagnetism. The metallic sample gives rise to a $B$-dependent background in the Hall resistance, making it challenging to observe the magnetic hysteresis loop. However, the $I$-induced hysteresis is clearly detectable in both $R_{\parallel}$ and $R_{\perp}$, since the Hall resistance background is unaffected by the current bias. Both the magnetic and current-induced hysteresis exhibits a sharp onset with decreasing temperature (Fig.~\ref{fig4}c and Fig.~\ref{figBloopT}), which defines the emergence of the orbital ferromagnetic order. 

The area of the hysteresis loop, $A_{\perp}^{B}$ and $A_{\perp}^{I}$, offers a gauge for the strength of the orbital ferromagnetic order. With decreasing temperature,  the sharp onset in $A_{\perp}^{B}$ and $A_{\perp}^{I}$ (top panel of Fig.~\ref{fig4}d) coincides with the onset of transport nonreciprocity, which is shown as the temperature dependence of  $V_1$ and $V_3$ (bottom panel of Fig.~\ref{fig4}d), extracted from the angular oscillation of nonreciprocity. The temperature dependence points towards a direct link between transport nonreciprocity, valley polarization, and the orbital ferromagnetism.

It should be noted that valley polarization breaks $C_2$ and $T$, but it preserves their product $C_2T$. Since the realization of an orbital ferromagnetic order requires $C_2T$ breaking, valley polarization in tTLG is distinct from orbital ferromagnetism. 
We propose that $C_2T$ breaking is not a necessary ingredient for transport nonreciprocity.  This is further supported by the fact that nonreciprocity is observed universally across the moir\'e band (Fig.~\ref{NonreciprocityMoire}), even in the absence of the anomalous Hall effect. The strength of nonreciprocity, reflected by the angular oscillation amplitude $V_1$, exhibits a density-modulation with the cascade of Fermi surface reconstructions (see Fig.~\ref{NonreciprocityMoire}a). This highlights the direct link between nonreciprocity and a valley-imbalanced charge carrier occupation. 

Fig.~\ref{fig4}e shows schematic diagram of valley-imbalanced Fermi surface occupation, which includes scenarios with three Fermi pockets (Panel \textit{i} and \textit{ii}) or one large Fermi surface per valley (Panel \textit{iii} and \textit{iv}). A valley-imbalance could also arise from a partial polarization, where carriers occupy both valleys (Panel \textit{ii} and \textit{iv}). Regardless of the detail, a valley-imbalanced Fermi surface occupation has three mirror axes (black dashed lines in Fig.~\ref{fig4}e). The angular dependence of nonreciprocity with non-zero $V_1$,  therefore, requires an additional $C_3$ symmetry breaking ingredient. 
This symmetry requirement is naturally satisfied by the exchange-driven instability in the momentum space, which breaks $C_3$ by inducing charge carrier condensation into a sub-set of Fermi pockets ~\cite{Dong2021momentum,Jung2015momentum,Huang2022momentum,Cheung2018momentum}. Fig.~\ref{fig4}f shows a series of Fermi surface occupations with one mirror axis (black dashed line). 
Panel \textit{i} displays a state with full momentum polarization, where all carriers condense into one Fermi pocket. Whereas panel \textit{iv} represents a partial momentum polarization, with one Fermi pocket being slightly larger compared to the other five.  It should be noted that nonreciprocity diminishes for a valley balanced carrier occupation. This allows valley and momentum polarization to be detected through nonreciprocity with high sensitivity. As a result, the observed nonreciprocity could arise from an extremely subtle imbalance in carrier occupation across Fermi pockets and valleys. This allows the momentum-space instability to emerge as a perturbation to other types of electronic order, such as orthorhombic anisotropy, while still accounts for the observed angular dependence in nonreciprocity.

Next, we discuss the potential influence of certain scattering mechanism, such as skew scattering and Berry curvature dipole ~\cite{Ma2019nonlinear,Kang2019nonlinear,He2022nonlinear,Sinha2022nonlinear}. Apart from the hysteretic rotation in the mirror axis, we note that nonreciprocity in tTLG is observed only at $T < 5$ K (Fig.~\ref{fig4}d and Fig.~\ref{figSIT}), which is much lower compared to the typical onset temperature of  scattering around hundreds of Kelvin ~\cite{Ma2019nonlinear,Kang2019nonlinear,He2022nonlinear,Sinha2022nonlinear}. Nevertheless, the presence of scattering could influence transport nonreciprocity indirectly. The exact role of scattering will certainly attract future efforts aiming to better understand the microscopic origin of the transport nonreciprocity.

Despite the lack of a microscopic understanding, it is instructive to examine the phenomenological behavior of the superconducting diode effect. Nonreciprocity in the superconducting transport is defined as an inequality in the critical supercurrent between forward and reversed current bias, $\eta_{I_c}(\phi)=I_c(\phi)-I_c(\phi+\pi)$  ~\cite{Lin2022SDE,DiodeTheoryPaper}. To minimize the potential influence of the normal state at large current, we operationally define the critical current $I_c$ as the current bias where the measured differential resistance becomes larger than the noise floor ~\cite{Benyamini2019fragility}, which is marked by vertical arrows in Fig.~\ref{EtaIc}. At the optimal density of the superconducting phase, the angular dependence of $\eta_{I_c}$, shown as the bottom left inset in Fig.~\ref{fig4}g, is best fit by a cosine function with a period of $2\pi$. This defines a single mirror axis, which indicates $C_3$ symmetry breaking. Since the superconducting diode effect also requires simultaneously breaking $C_2$ and $T$, the angular dependence of superconducting nonreciprocity points towards the same symmetry requirement as the momentum-space instability.

Fig.~\ref{fig4}g-h plots the angular dependence of nonreciprocity measured from the normal state at $T > T_c$ (Fig.~\ref{fig4}g) and $I > I_c$ (Fig.~\ref{fig4}h). The angular dependence points towards a single mirror axis that breaks $C_2$ and $C_3$ symmetry simultaneously. The similarity between the normal and superconducting phases raises the possibility that superconductivity inherits the broken symmetries of the normal state. In this scenario, the momentum-space instability offers a mechanistic explanation for the zero-field superconducting diode effect ~\cite{Lin2022SDE}.



\bibliography{Li_ref}

\newpage

\newpage
\clearpage

\section*{Method}

\renewcommand{\thefigure}{M\arabic{figure}}
\def\theequation{M\arabic{equation}}
\def\thetable{M\Roman{table}}
\setcounter{figure}{0}
\setcounter{equation}{0}

\begin{figure}[!b]
\includegraphics[width=0.85\linewidth]{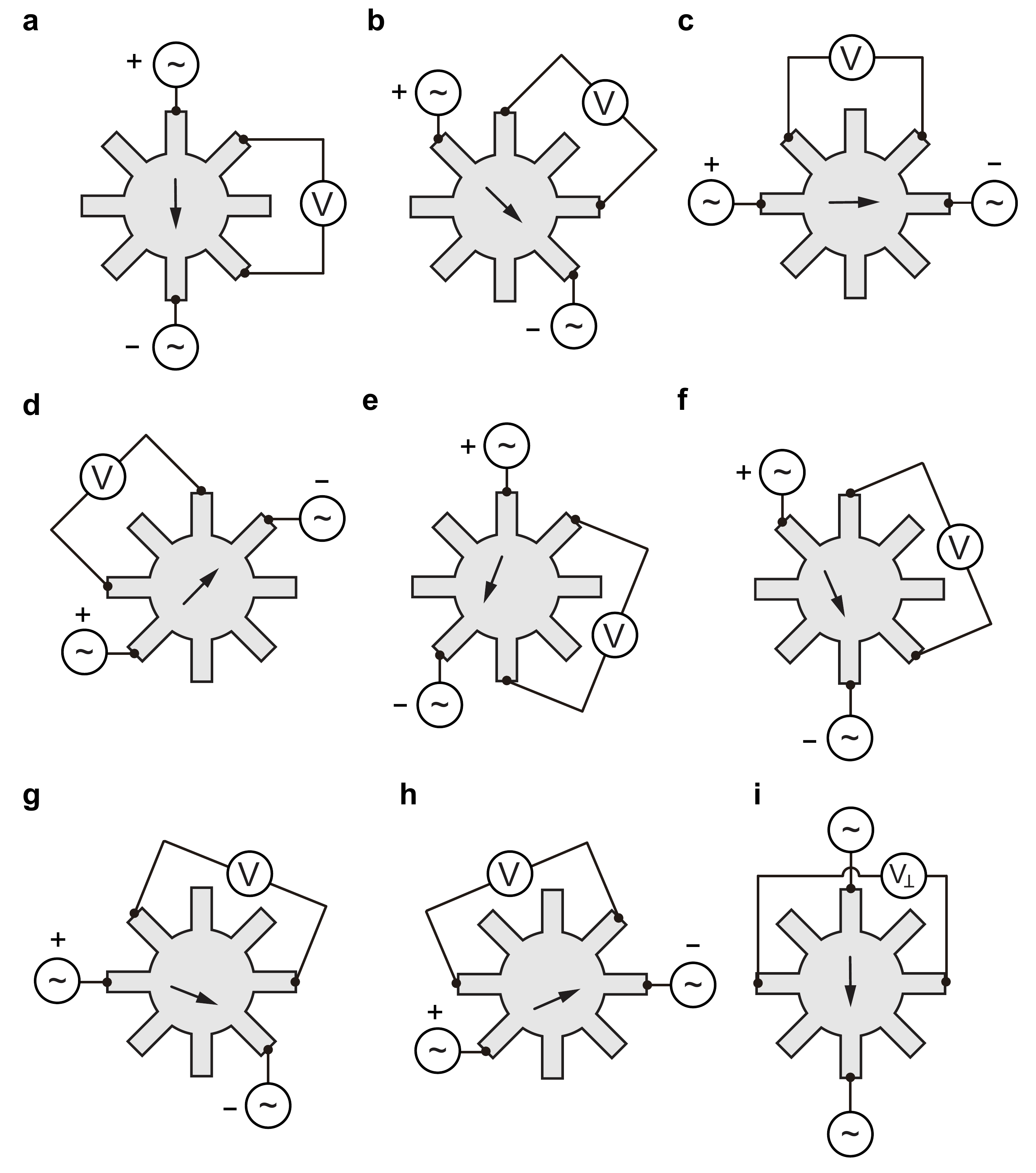}
\caption{\label{figSetup} {\bf{Measurement setup.} }   (a-h) Schematic diagram showing the measurement configuration with current flowing is different azimuth directions $\phi$ across the ``sunflower'' sample. For each $\phi$, the longitudinal transport response is defined as the voltage difference across two contacts (each contact resembles a petal of the ``sunflower''), which are aligned parallel to the direction of current flow. Panels (a) to (h) show measurement configurations for $8$ azimuth directions. Along the same vein, the transverse response is measured across two contacts aligned perpendicular to the current flow direction, as shown in panel (i). We use $V_{\parallel}$ ($V_{\perp}$) to denote the voltage difference across two contacts that are parallel (perpendicular) to the current flow direction (Fig.~\ref{fig1}a). Instead of applying current bias at the source contact and short to ground at the drain contact, we apply a positive current bias to the source contact, and a negative bias to the drain contact. This ensures that the center of the sample remains at zero electric potential, thus suppressing the potential influence of capacitive coupling and thermal-electric effects associated with the contact resistance. 
}
\end{figure} 

\subsection{Hysteresis in the Mirror Axis Orientation}

Apart from thermal cycle, hysteretic rotation in the mirror axis can also be induced by a large current bias. 

\begin{figure*}
\includegraphics[width=1\linewidth]{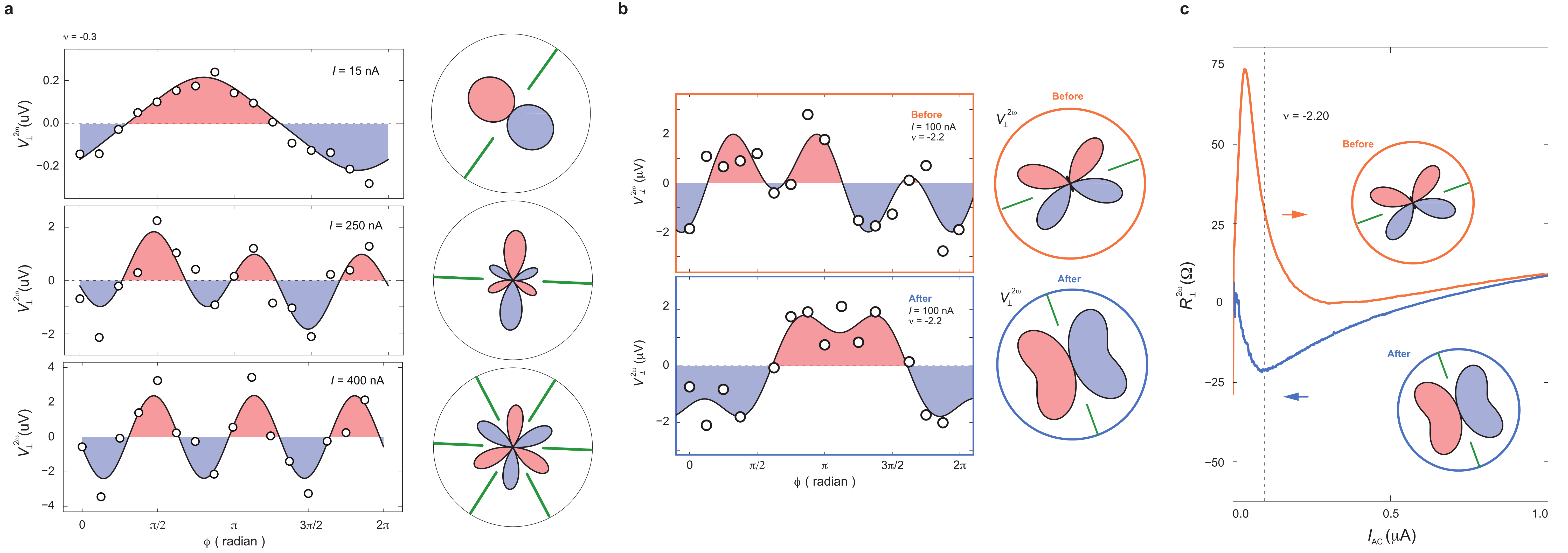}
\caption{\label{CurrentHysteresis} {\bf{Current-induced Hysteresis in nonreciprocity. }} (a) The angular dependence of nonreciprocity measured at the same moir\'e band filling $\nu=-0.3$ and different current bias. At a small current bias, the angular dependence is best described by a one-fold symmetric cosine function. With increasing current, the angular dependence starts to evolves into a mixture between one- and three-fold at $I >  200$ nA. Further increasing the current bias gives rise to angular dependence at $I = 400$ nA that is best captured by $\cos3\phi$. (b) The angular dependence of nonreciprocity at $\nu = -2.2$ before (top panel) and after (bottom panel) the application of a large current bias. The large current bias induces a hysteretic rotation in the underlying mirror axis, which is marked by the green solid line in the polar coordinate plots. (c) $R_{\perp}^{2\omega}$, defined as \Vto $/I$, as a function of current bias. As the current bias is swept back and forth, $R_{\perp}^{2\omega}$ exhibits a hysteresis loop. The angular dependence of \Vto\ in panel (b) is measured before and after this hysteresis loop at a fixed current bias of $I = 100$ nA.
Inset shows the schematic angular dependence for the data shown in panel (b). 
}
\end{figure*}

Fig.~\ref{CurrentHysteresis} demonstrates the current-induced tunability in the angular dependence of nonreciprocity. Fig.~\ref{CurrentHysteresis}a plots the angular dependence of \Vto\ measured at the same carrier density with different current bias. The angular dependence behaves like a cosine function with a period of $2\pi$ at low current bias. It changes to a linear combination of $\cos\phi$ and $\cos3\phi$ as current bias is increased to $I > 200$ nA. At $I = 400$ nA, the angular dependence becomes largely three-fold symmetric, albeit the quality of the angular fit becomes poorer at this large current.

The current-induced change in the angular dependence of nonreciprocity can be hysteretic, \emph{i.e.} the angular dependence of nonreciprocity, measured with a small current bias, is different before and after a large current bias. The hysteretic behavior is shown in Fig.~\ref{CurrentHysteresis}b. Before (after) applying a large current bias at $1 \mu$A, the angular dependence measured at $\nu = -2.2$ is highlighted by the orange (blue) box. In both cases, the angular dependence is a linear combination of  $\cos\phi$ and $\cos3\phi$. However, the mirror axis, marked by green solid lines, exhibits a prominent rotation before and after the current bias.

The hysteretic rotation gives rise to a hysteresis loop in the second harmonic nonlinear response $R_{\perp}^{2\omega}$ (Fig.~\ref{CurrentHysteresis}c), which is defined as \Vto $/I$. The insets in Fig.~\ref{CurrentHysteresis}c are schematic of the angular dependence measured before and after sweeping the current up and down. 

\begin{figure*}
\includegraphics[width=0.7\linewidth]{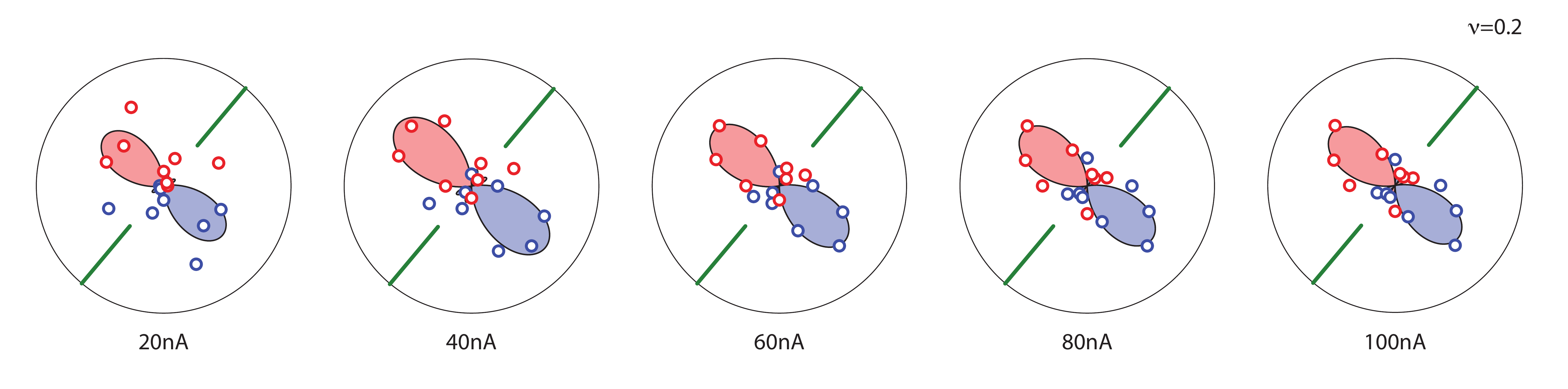}
\caption{\label{figSIcurrent} {\bf{Angle dependence of $\eta_{\parallel}$ measured at different DC current bias.}} Polar-coordinate plot of the angle dependence of $\eta_R$, measured at $B=0$, $T = 20$ mK and $\nu=0.2$. With increasing DC current bias, a similar angular dependence that is predominantly one-fold symmetric is observed up to $I_{DC} = 100$ nA, all pointing towards around $140^\circ$. 
}
\end{figure*}

Furthermore, we note that current-induced hysteresis usually occurs at large current bias. Fig.~\ref{figSIcurrent} shows the angular dependence of nonreciprocity $\eta_{\parallel}$ measured with different DC current bias. Up to $I = 100$ nA, the angular dependence of nonreciprocity, along with its associated mirror axis orientation, remain the same. As such, the angular dependence reported in this work is mostly carried out at $I \leq 100$ nA, unless otherwise specified.

\subsection{Nonreciprocity and nonlinearity}

\begin{figure}
\includegraphics[width=1\linewidth]{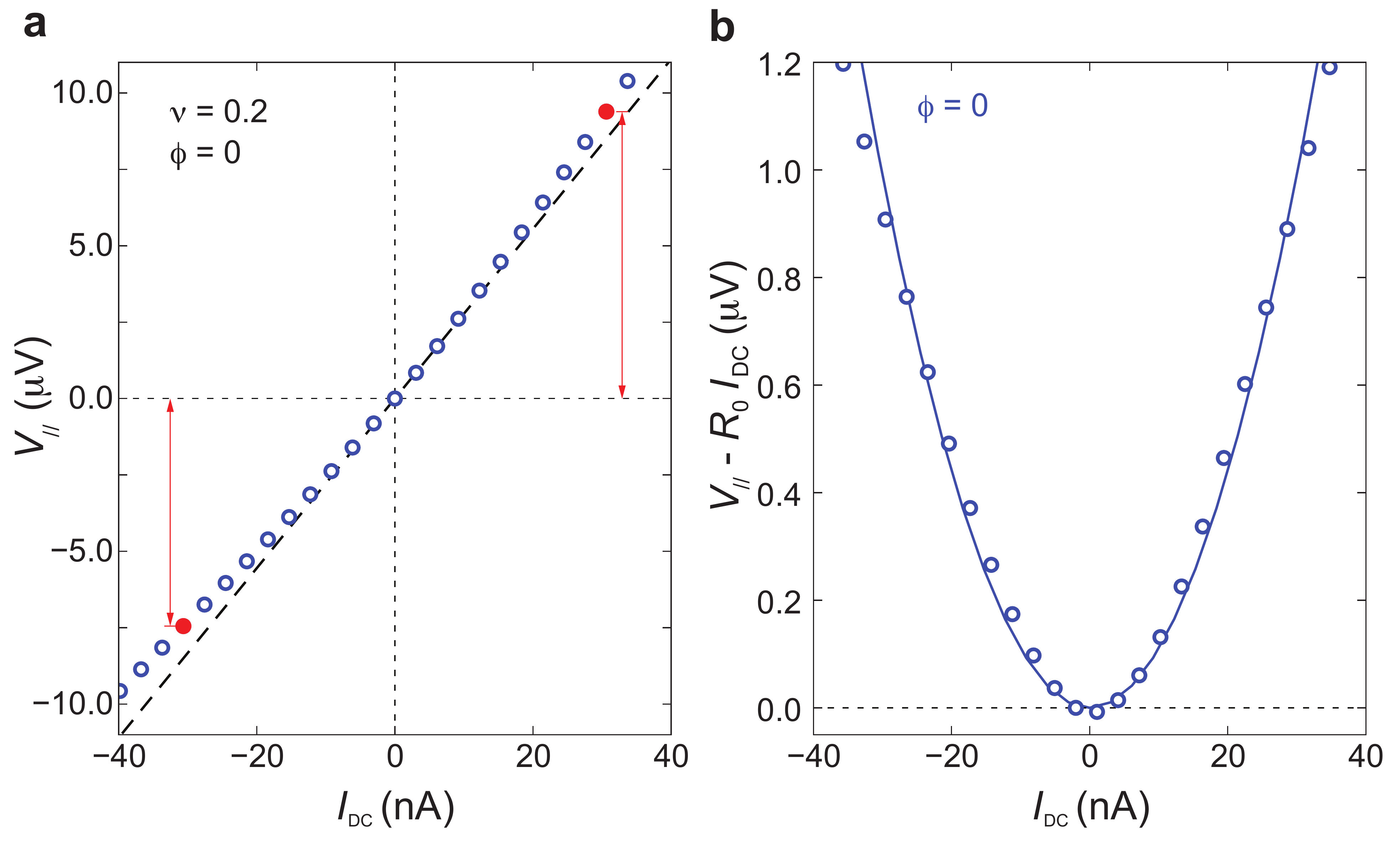}
\caption{\label{DCEta} {\bf{Nonreciprocity and nonlinearity. }} (a) Current-voltage characteristic measured at $\nu=-0.3$ with DC current flowing along azimuth angle of $\phi = 0^{\circ}$. The black dashed line is a linear fit to the portion of the IV curve near zero current bias. (b) The nonreciprocal component of the IV curve, which is extracted by subtracting the ohmic component of the transport response, $\eta = V_{\parallel}-I_{\textrm{DC}}R_0$. $R_0$ denotes the slope of the IV curve at $I_{DC}=0$ The solid line is a quadratic fit to the current dependence of $\eta$. 
}
\end{figure}

\begin{figure}[h]
\includegraphics[width=0.85\linewidth]{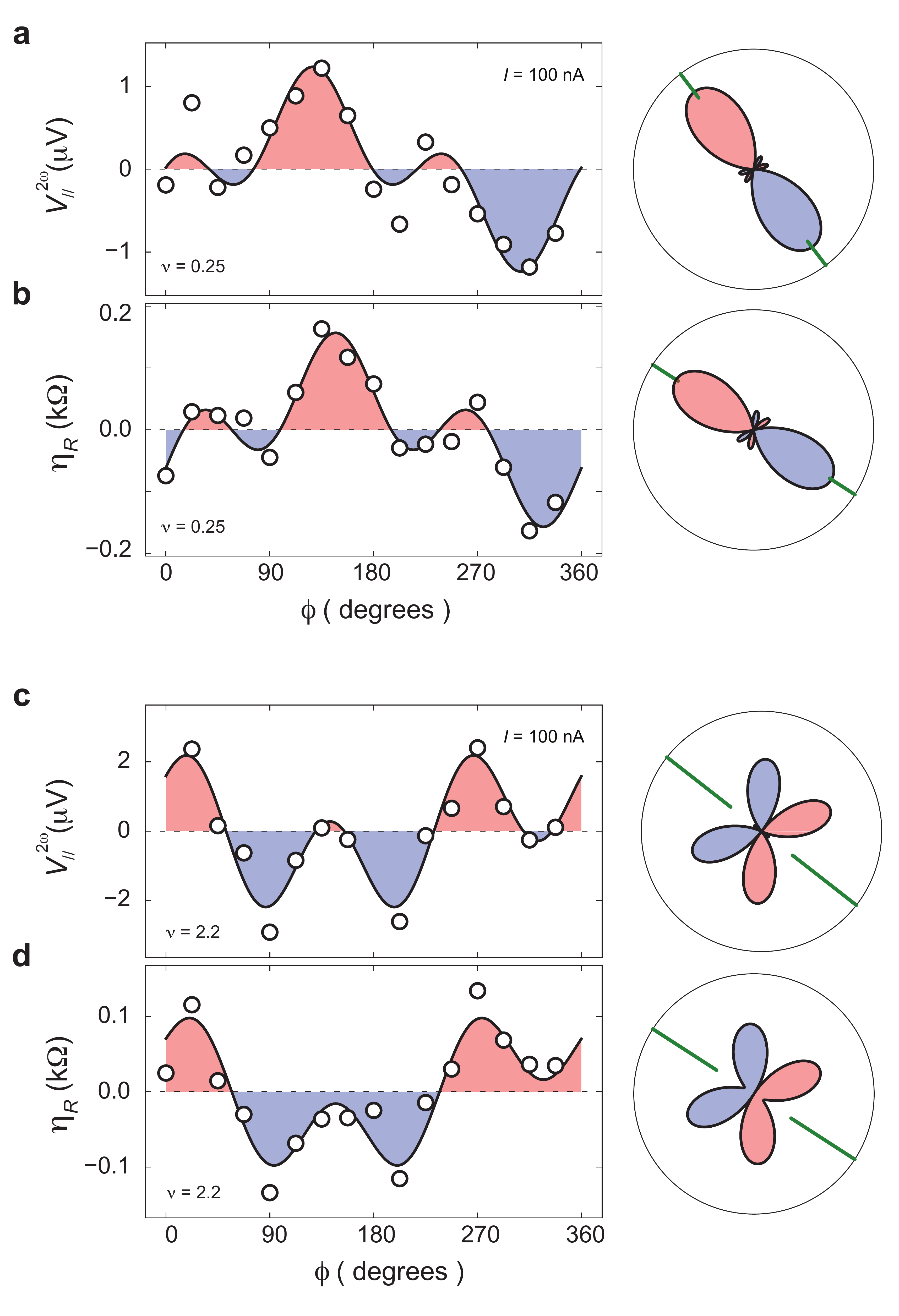}
\caption{\label{ACDCcompare} {\bf{Second-harmonic nonlinear response and transport nonreciprocity. }} 
Angle dependence measurement of (a) \Vpo\ and (b) $\eta_R$ at $\nu=0.25$ and (c) \Vpo\ and (d) $\eta_R$ at $\nu=2.2$. The remarkable match in the angle dependence between \Vpo\ and (b) $\eta_R$ at different densities illustrates their correspondence relation. \Vpo\ are measured at $I_{AC}=100$ nA, $\eta_R$ are measured at $I_{DC}=100$ nA. All measurements are performed at $B=0$ and $T=20$ mK. 
}
\end{figure}

Transport nonreciprocity $\eta$ can be extracted from the I-V curve measured with a DC current bias.  As shown in Fig.~\ref{DCEta}a, the IV curve deviates from the linear response (black dashed line) at large current bias. As a result of the deviation, longitudinal resistance $R_{\parallel}$, defined as $R_{\parallel} = V_{\parallel}/I$, is larger at the positive current bias compared to the negative bias (red circles in Fig.~\ref{DCEta}a). Subtracting the linear component reveals the nonreciprocal response, $\Delta V_{\parallel} - R_0 I_{DC}$, where $R_0$ denotes the slope of the IV curve at $I_{DC}=0$. The nonreciprocal component is shown to have a quadratic current dependence, $\eta \sim I^2$ (the black solid line in Fig.~\ref{DCEta}b is a quadratic fit to the data). 

Given the quadratic current dependence in $\eta$, nonreciprocity can be expressed as $\sim sin^2{\omega t} \sim sin(2\omega t)$ in the presence of an AC current bias with frequency $\omega$. As such, $\eta$ is equivalent to the nonlinear transport response measured at the second harmonic frequency of an AC current bias.

Fig.~\ref{DCEta}c-d shows the a.c. current dependence of the second-harmonic nonlinear response in the longitudinal and transverse channel, \Vpo\ and \Vto. The nonlinear response in both channels exhibit a sign reversal upon reversing the current flow direction, even though the measurement configurations respect chiral symmetry. Such a sign reversal provides a strong indication for $C_2$ symmetry breaking.

Fig.~\ref{ACDCcompare} compares the angular dependence of transport nonreciprocity with the nonlinear response at the second harmonic frequency. The azimuth angle $\phi$ defines the direction of current flow across the sunflower-shaped sample. Different $\phi$ is realized using the measurement configurations shown in Fig.~\ref{figSetup}.  

Fig.~\ref{ACDCcompare}b and d plots the angular dependence of $\eta$, which are extracted from d.c. transport IV curves with current flowing in different azimuth directions. The value of $\eta$ along $\phi$ is defined as the difference in $R_{\parallel}$ between current flowing in the azimuth direction of $\phi$ and $\phi + \pi$, 
\begin{equation}
\eta_{\parallel}(\phi)=R_{\parallel}(\phi)-R_{\parallel}(\phi+\pi).
\end{equation}
\noindent Here $R_{\parallel}$ is the transport response at a fixed d.c. current value $I_{DC} = 100$ nA. By definition, $\eta$ displays a sign reversal between forward and reversed current bias.

Fig.~\ref{ACDCcompare}a and c plots the angular dependence of the second-harmonic nonlinear transport response $V^{2\omega}$, which is measured at the same band fillings as panel b and d. The value of $V^{2\omega}$ along $\phi$ is simply defined as the voltage response at the second-harmonic frequency as an a.c. current is biased along azimuth direction of $\phi$. Based on this comparison, we make a few important observations: (i) the angular dependence of $\eta$ shows excellent agreement with that of the second-harmonic nonlinear response $V^{2\omega}$; (ii) the angular dependence can be captured by a linear combination of one-fold and three-fold symmetric components; (iii) the one-fold symmetric component of the angular dependence is aligned along the corner of the Brillouin zone. The azimuth directions of all six corners are marked by dashed lines in the polar-coordinate plots of  Fig.~\ref{ACDCcompare}. Combined, the angular dependence of transport nonreciprocity and the second-harmonic nonlinear response offers further confirmation for the exchange-driven instability in the valley and momentum space.

\begin{figure*}
\includegraphics[width=0.73\linewidth]{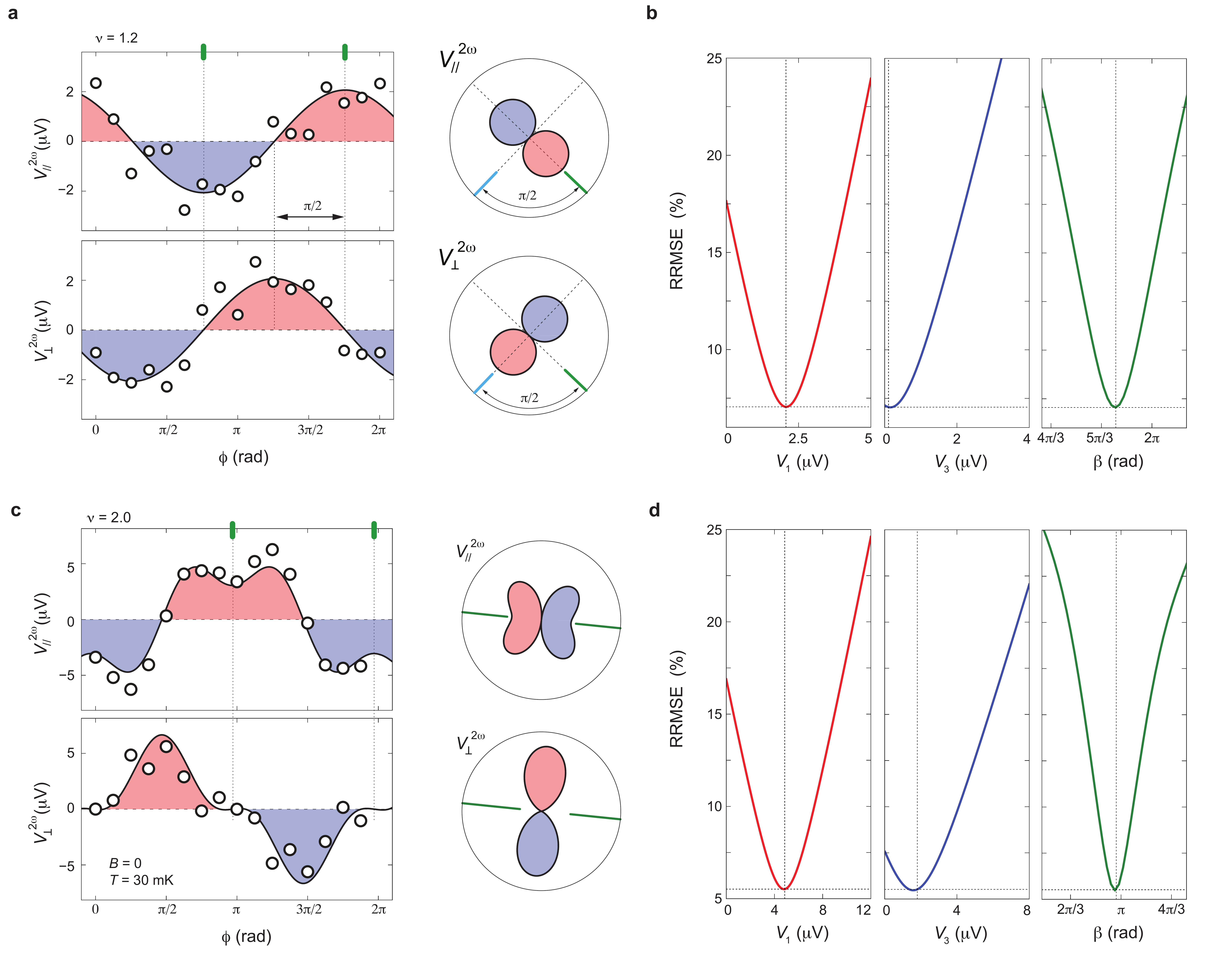}
\caption{\label{figError} {\bf{Relative Root Mean Squared Error in the angular fit with large $V_1$.} }  (a) The angular dependence of nonreciprocity measured at $\nu = 2.15$. (b) Relative root-mean-squared error (RRMSE), defined according to Eq.~M2, as a function of $V_1$ (left panel), $V_3$ (middle panel), and $\beta$ (right panel). (c) The angular dependence of nonreciprocity measured at $\nu = 2.15$. (d) Relative root-mean-squared error (RRMSE), defined according to Eq.~M2, as a function of $V_1$ (left panel), $V_3$ (middle panel), and $\beta$ (right panel). 
}
\end{figure*}

\begin{figure*}
\includegraphics[width=0.73\linewidth]{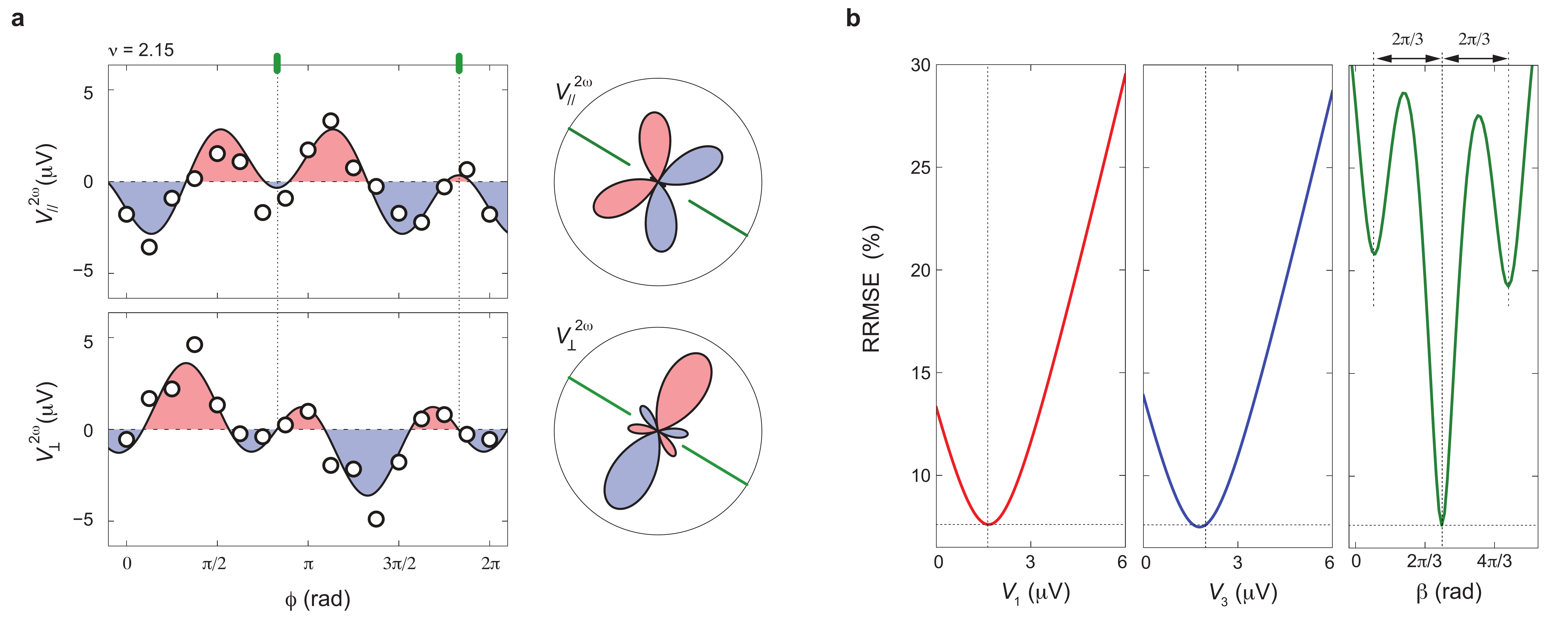}
\caption{\label{figError3fold} {\bf{Relative Root Mean Squared Error in the angular fit with large $V_3$.} }  (a) The angular dependence of nonreciprocity measured at $\nu = 2.15$. (b) Relative root-mean-squared error (RRMSE), defined according to Eq.~M2, as a function of $V_1$ (left panel), $V_3$ (middle panel), and $\beta$ (right panel). (c) The same angular dependence results as shown in panel (a), buth with a different fit function with $V_1 = 0 $. 
}
\end{figure*}

\begin{figure*}
\includegraphics[width=0.87\linewidth]{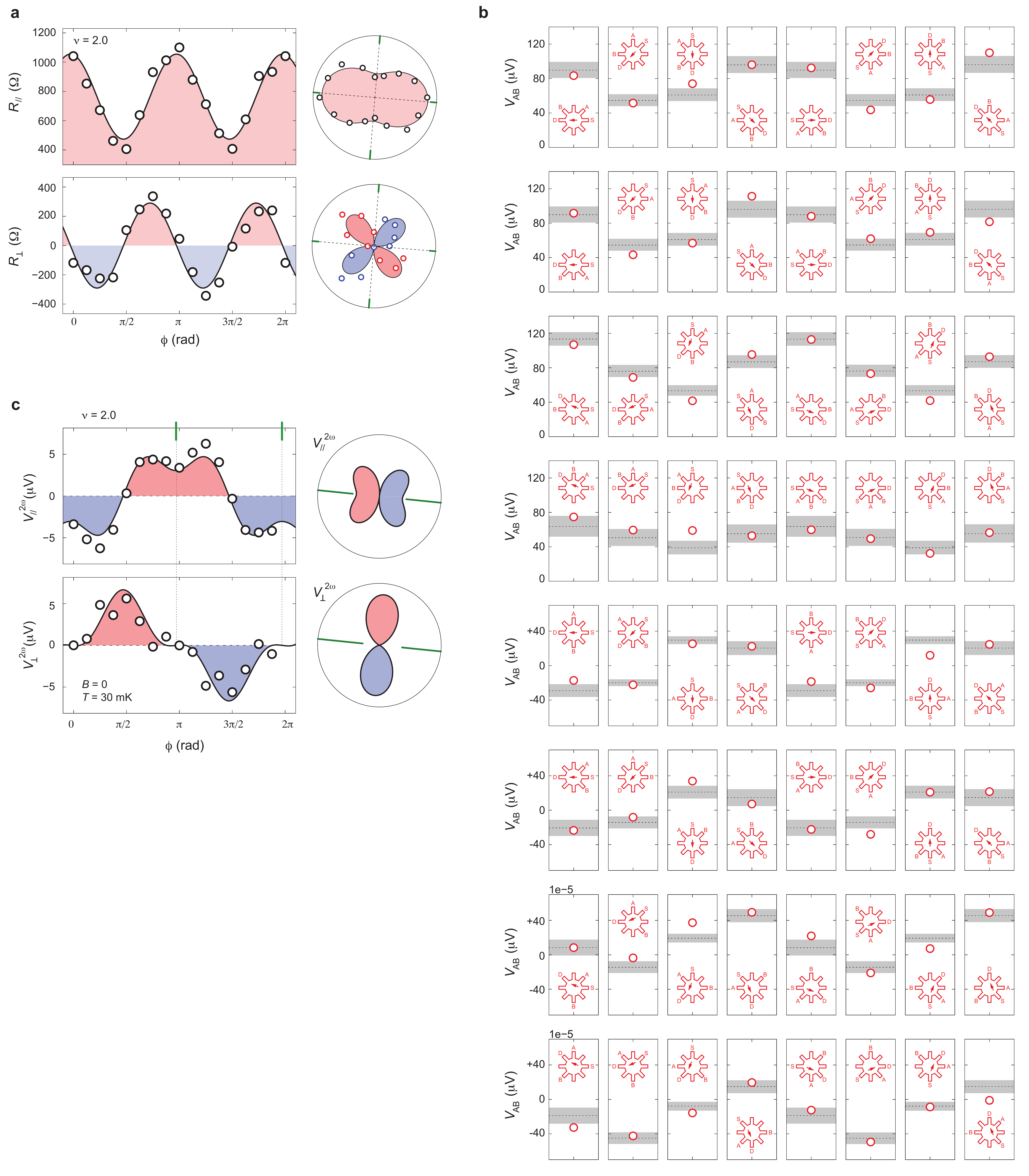}
\caption{\label{Fullfit} {\bf{The ``sunflower'' model beyond angular fit.} }  (a) The angular dependence of longitudinal and transverse resistance, $R_{\parallel}$ and $R_{\perp}$, measured at $\nu = 2$.  (b) Measurements from different ``sunflower'' configurations compared to the expected value extracted from a single conductivity matrix ~\cite{Vafek2022sunflower,Zhang2022sunflower}. The RRMSE of the fit is $1.74\%$. (c) The angular dependence of longitudinal and transverse nonreciprocity, $\eta_{\parallel}$ and $\eta_{\perp}$, measured with an AC current of $100$ nA at $\nu = 2$. 
}
\end{figure*}

\subsection{Mean-square Error in the angular fit}

The procedure for determining the best angular fit for transport nonreciprocity is demonstrated in Fig.~\ref{figError}. In general, the best angular fit is obtained by minimizing the relative root-mean-squared error (RRMSE), which is defined as
\begin{equation}
RRMSE=\frac{\sqrt{\frac{1}{N}\sum_i (V_{i}-V^0_i)^2}}{\sqrt{\sum_i (V_{i})^2}},
\end{equation}
\noindent where $V_i$ is the $i$th point of the measurement results and $V_i^0$ is the value of the best fit at the corresponding azimuth direction. $N$ denotes the number of points. $N=32$ for the angular fit for longitudinal and transverse responses, since there are $16$ angles for each transport channel.
Fig.~\ref{figError}b and d shows the evolution of the RRMSE as one of the fit parameters deviates from the parameter of the best angular fit, while the other two fit parameters are fixed at the value of the best fit. According to Eq.~1, the fit parameter $V_1$, $V_3$, and $\beta$ creates a three-dimensional phase space for identifying the best angular fit.  For the angular dependence in Fig.~\ref{figError} a and c, RRMSE is around $5 \%$ for the best fit. This indicates that the angular dependence of nonreciprocity is well-captured by Eq.~1. 

Fig.~\ref{figError3fold} examines the best fit for the angular dependence at $\nu = 2.15$, where the best fit features a strong 3-fold symmetric component. The RRMSE of the best fit is around $8\%$. While slightly larger compared to Fig.~\ref{figError}, a RRMSE below $10\%$ is considered a good angular fit. That a predominantly three-fold symmetric behavior has slightly higher error is understandable, since the angular oscillation  is approaching the resolution of the sunflower geometry. We note that the right panel of Fig.~\ref{figError3fold}b exhibits two extra local minima in RRMSE, which are located $2\pi/3$ away from the optimal value of $\beta$. These two local minima provides a strong indication for the three-fold symmetric component in the angular dependence of nonreciprocity.

\begin{figure*}
\includegraphics[width=0.77\linewidth]{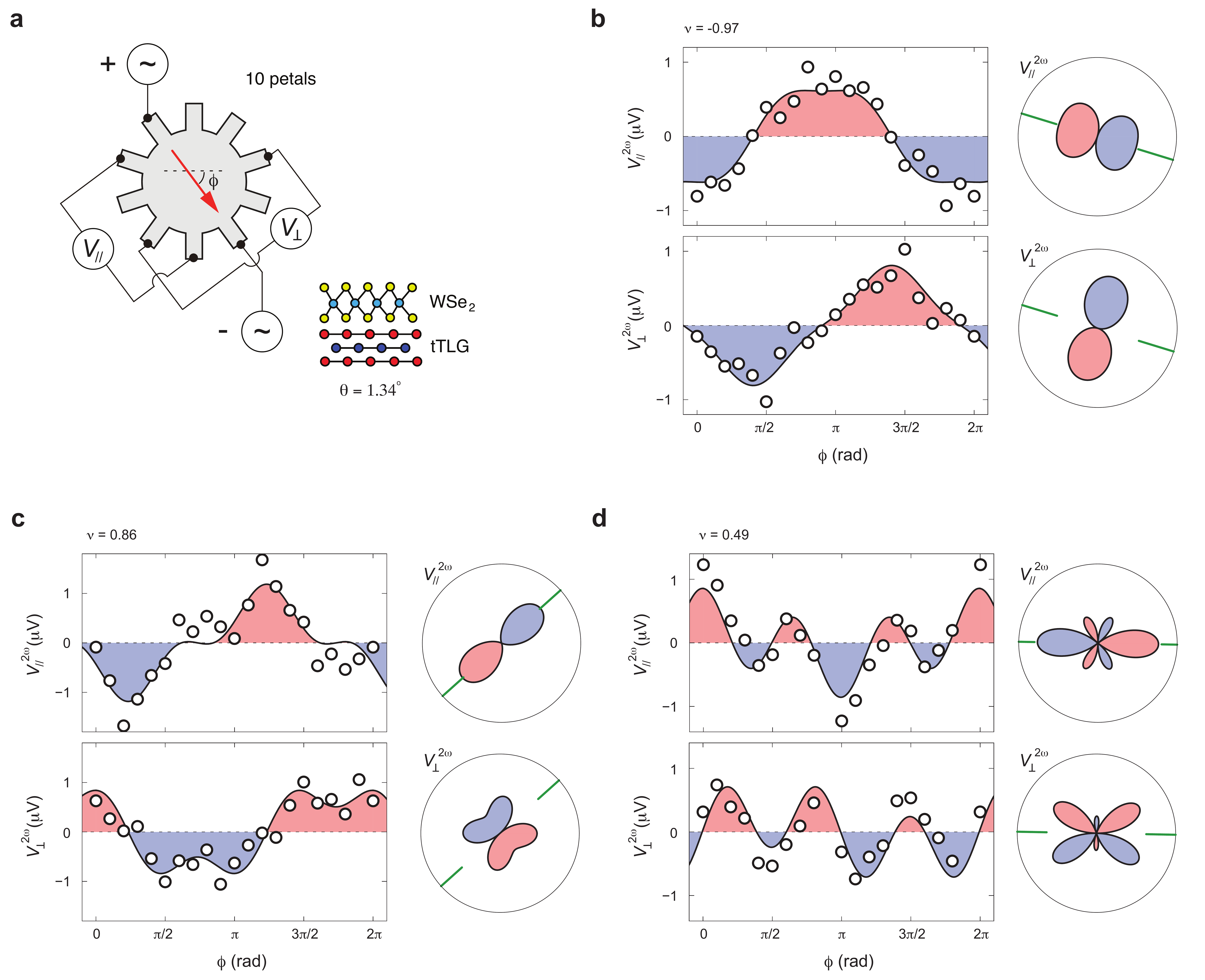}
\caption{\label{fig10petal} {\bf{Angle-resolved nonreciprocity measurement with higher angular resolution.}}  (a) Schematic diagram of the ``sunflower''-shaped sample with 10 petals. The increased number of electrical contacts enables higher angular resolution in the nonreciprocity measurement.  (b-d) Angle-resolved nonreciprocity measured at different moir\'e band filling of the tTLG sample, which has a twist angle of $\theta = 1.34^{\circ}$. The best angular fit for the angular dependence (black solid line) is captured by Eq.~1, which is extracted by minimizing the RRMSE of nonreciprocity from both longitudinal and transverse channels. With increased angular resolution, the best fit to the angular dependence of nonreciprocity in panel (b) features RRMSE of $0.3\%$, while other angular dependence shows RRMSE of less than $2\%$. Beyond the one-fold and three-fold components, the next lowest order angular component is described by $\cos(5\phi)$ and $\sin(5\phi)$. While an oscillatory period of $72^{\circ}$ is well within the measurement resolution, such a component is not observed in our measurement. This is a strong indication that the angular components with $N>3$ has small oscillatory amplitude. All measurement performed at $T=20$ mK, $B=0$, and $I_{AC}=50$ nA.
}
\end{figure*}

\subsection{The potential impact of moir\'e disorder, sample uniformity and ballistic transport}

Fig.~\ref{Fullfit}a plots the angular dependence of the linear transport response, $R_{\parallel}$ and $R_{\perp}$. The angular dependence is in excellent agreement with the expected behavior of an orthorhombic anisotropy, as shown in Fig.~\ref{fig1}d. The angular oscillation points towards two orthogonal mirror axes. The sample exhibits different conductivity as current flows along each mirror axis. The angular dependence in Fig.~\ref{Fullfit}a can be captured by a conductivity matrix. According to Ref.~\cite{Vafek2022sunflower}, knowledge of the conductivity matrix allows us to compute the potential distribution across the entire sample, as long as electrons move through the sample diffusively, \emph{i.e.} electron transport is non-ballistic. 

The ``sunflower'' sample geometry features eight electrical contacts (petals) attached to the circumference of a circular sample channel. The geometry allows more than $400$ independent measurement configurations: picking any two contacts as source and drain, a voltage measurement can be performed across any two of the remaining contacts. While $400$ configurations appear redundant, they allow us to fully map the potential distribution across the circumference of the sample. By fitting these measurements with a single conductivity matrix with the ``sunflower'' model, we can gain insight into the uniformity of transport response across the sample.

Fig.~\ref{Fullfit}b plots results from a sub-set of all independent measurement configurations. Red circles denote the measured value, insets show schematic diagrams of each measurement configuration. The horizontal gray stripe in Fig.~\ref{Fullfit}b is the expected value predicted by the ``sunflower'' model based on a single conductivity matrix ~\cite{Vafek2022sunflower}. The width of the gray stripe  arises from the non-zero width of each electrical contact, which gives rise to a range of possible electrical potential.

Noticeably, measurement configurations from the same row  are associated with an in-plane rotation. In the absence of electronic anisotropy, we anticipate the same measurement results from different configurations of the same row. That is to say, variations in the transport response across each row arises directly from the presence of electronic anisotropy. Different measurement configurations probe slightly different parts of the sample. For example, configurations in the fourth row from the top probe the transport response close to the circumference the sample. Results from all measurement configurations are reasonably captured by a single conductivity matrix. The RRMSE of this fit is around $0.017$, which points towards an excellent agreement between the measurement and the model. The excellent fit provides a strong indication that the transport response is uniform across the entire sample. 

As a by-product, the fit in  Fig.~\ref{Fullfit}b suggests that electron transport through the sample in a non-ballistic. 

Fig.~\ref{Fullfit}c displays the angular dependence of transport nonreciprocity measured at the same band filling. Green solid lines mark the orientation of the mirror axis. At this band filling, the mirror axis extracted from nonreciprocity is close to being aligned with the principle axis of the linear component of the transport response. In the SI section titled ``Nematicity and momentum polarization'', we offer more detailed discussion regarding the relationship between the mirror axis of nonreciprocity and the principle axis of the linear transport.

\section*{Acknowledgments}
J.I.A.L. wishes to thank Oskar Vafek and Dima Feldman for stimulating discussions. L.F. thanks the organizers of the workshop ``Superconducting diode effects" on Virtual Science Forum, which ignited the collaboration. 
N.J.Z. acknowledge support from the Jun-Qi fellowship. J.I.A.L. acknowledge funding from NSF DMR-2143384. Device fabrication was performed in the Institute for Molecular and Nanoscale Innovation at Brown University. D.V.C. acknowledges financial support from the National High Magnetic Field Laboratory through a Dirac Fellowship, which is funded by the National Science Foundation (Grant No. DMR-1644779) and the State of Florida.
K.W. and T.T. acknowledge support from the Elemental Strategy Initiative
conducted by the MEXT, Japan (Grant Number JPMXP0112101001) and  JSPS
KAKENHI (Grant Numbers 19H05790, 20H00354 and 21H05233). The work at Massachusetts Institute of Technology was supported by a Simons Investigator Award from the Simons Foundation.

\newpage

\newpage
\clearpage

\pagebreak
\begin{widetext}
\section{Supplementary Materials}

\begin{center}
\textbf{\large Angle-resolved transport nonreciprocity and spontaneous symmetry breaking in twisted trilayer graphene}\\
\vspace{10pt}

Naiyuan James Zhang,
Jiang-Xiazi Lin, Dmitry V. Chichinadze, 
Yibang Wang, Kenji Watanabe, Takashi Taniguchi, Liang Fu, and J.I.A. Li$^{\dag}$

\vspace{10pt}
$^{\dag}$ Corresponding author. Email: jia$\_$li @brown.edu
\end{center}

\noindent\textbf{This PDF file includes:}

\noindent{Figs. S1 to S13}

\noindent{Supplementary Text}

\noindent{Supplementary Data}

\noindent{Materials and Methods}

\renewcommand{\vec}[1]{\boldsymbol{#1}}

\renewcommand{\thefigure}{S\arabic{figure}}
\def\theequation{S\arabic{equation}}
\def\thetable{S\Roman{table}}
\setcounter{figure}{0}
\setcounter{equation}{0}

\section{Supplementary Text}

\subsection{Nematicity and momentum polarization}

There are several possible scenarios regarding the connection between momentum polarization and nematicity. 

First, an electronic order with rotational symmetry breaking could still preserves the two-fold inplane rotation symmetry $C_2$.  For example, a quadruple distortion in the Fermi surface will give rise to an orthorhombic anisotropy, but it does not generate a nonreciprocal response. Such a situation is demonstrated by Fig.~\ref{LinearNonlinear10K}. At high temperature $T = 10$ K, nonreciprocity is suppressed, whereas the orthorhombic anisotropy persists.

\begin{figure*}[!b]
\includegraphics[width=0.8\linewidth]{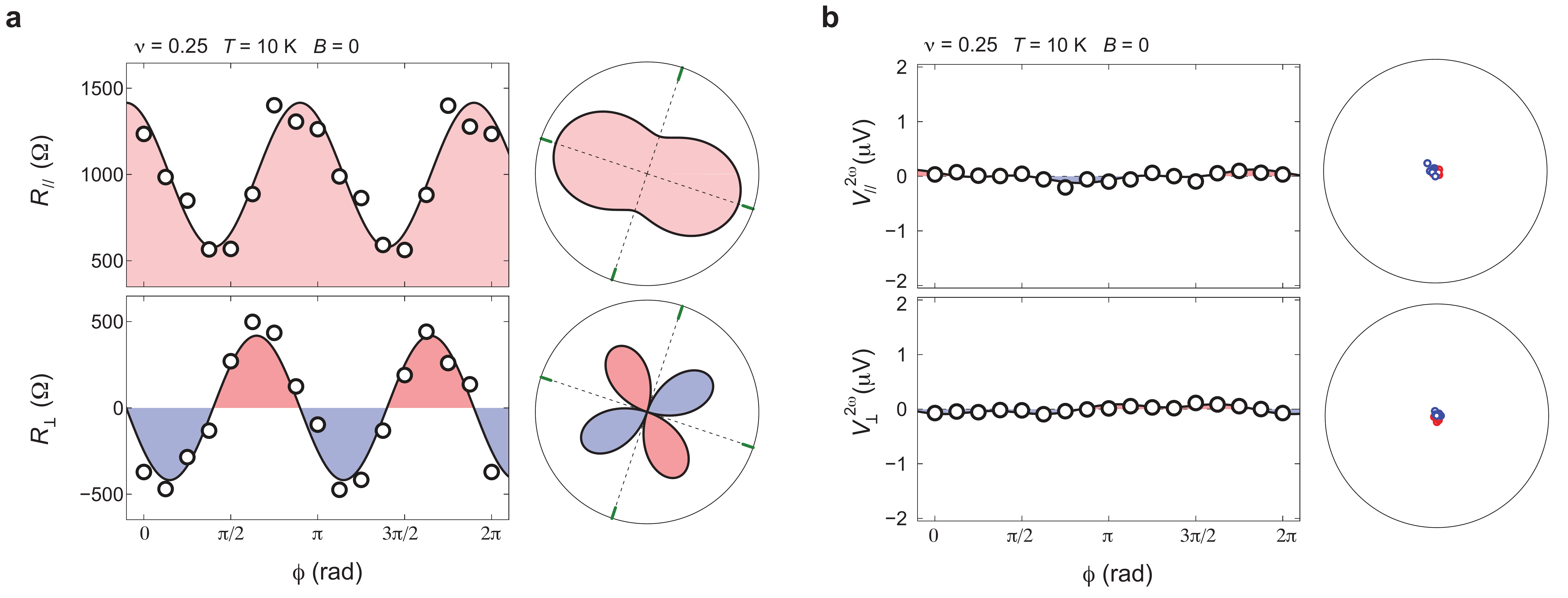}
\caption{\label{LinearNonlinear10K} {\bf{Anisotropic linear response and vanishing nonlinear transport.} }  (a) Linear transport response, $R_{\parallel}$ and $R_{\perp}$, and (b) nonreciprocal transport response, \Vpo\ and \Vto, as a function of azimuth angle of current flow $\phi$ measured at $\nu = 0.25$. All measurements are performed at $T = 10$ K, $B = 0$, and $I_{AC}=100$ nA. The vanishing nonreciprocal response at high temperature is consistent with the temperature dependence of nonreciprocity in Fig.~\ref{fig4}d and Fig.~\ref{figSIT}. 
}
\end{figure*} 

Secondly, the momentum space instability could be the only source of rotational symmetry breaking. In this case, the mirror axis of the nonreciprocal response would align with that of the linear transport. This scenario is consistent with the angular dependence of linear and nonlinear response in Fig.~\ref{Fullfit}a and c. 

While a momentum polarized Fermi surface occupation, as shown in Fig.~\ref{fig4}f, induces an angular oscillation in the sample conductivity, the oscillation amplitude $\delta \sigma$ may be much smaller compared to the overall sample conductivity $\sigma$. As such, the angular dependence in $\sigma$ is almost undetectable and the observed linear transport response is mostly isotropic. On the other hand, a $C_2$ symmetric order generates zero background in transport nonreciprocity. As a result, transport nonreciprocity is sensitive to an arbitrarily weak imbalance in the Fermi surface occupation across opposite valleys and different Fermi pockets.

Lastly, different types of Fermi surface distortions could occur simultaneously.  In the scenario where the quadruple distortion is more prominent, the momentum space instability can be treated as a perturbation, which is mostly independent of the orthorhombic anisotropy. In this scenario, the mirror axis of the orthorhombic anisotropy can be misaligned from that of momentum polarization. As a result, the angular dependence of the linear transport response could exhibits a different mirror axis orientation compared to that of nonreciprocity. This scenario offers a possible explanation for the data shown in Fig.~\ref{LinearNonlinearnu1}.

\begin{figure*}
\includegraphics[width=0.8\linewidth]{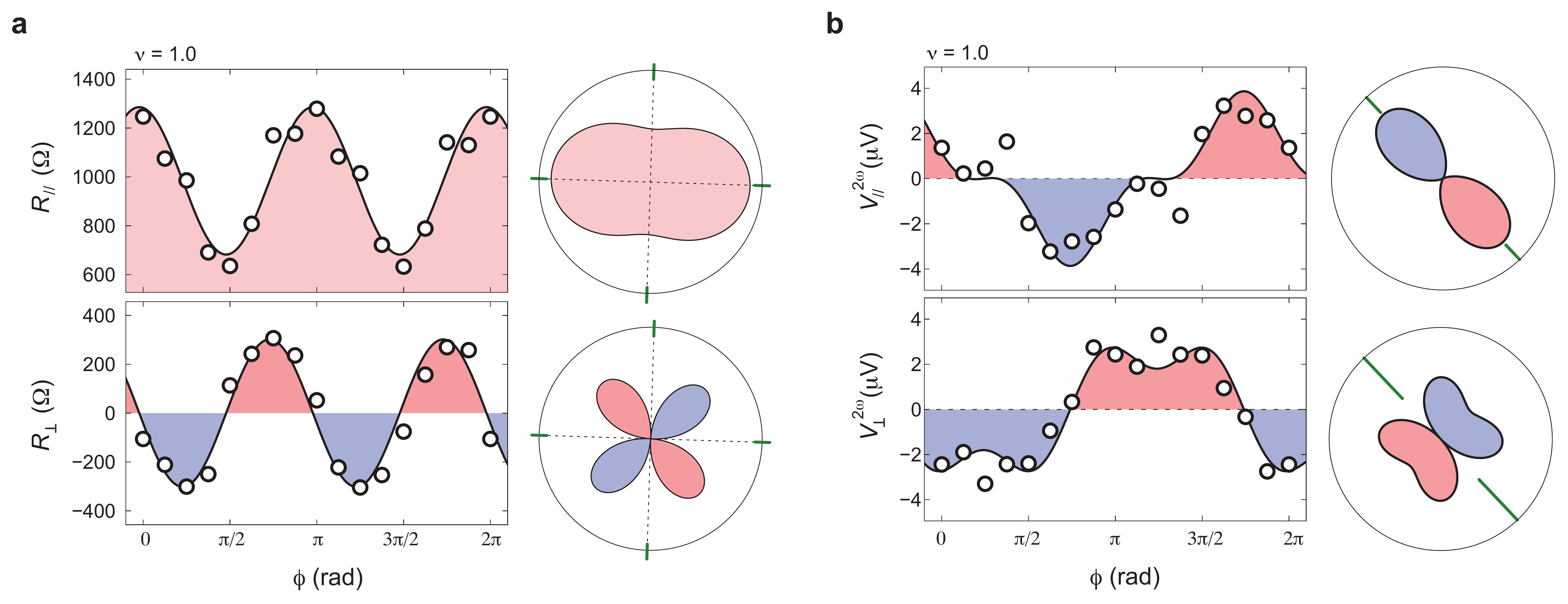}
\caption{\label{LinearNonlinearnu1} {\bf{Anisotropic linear response and nonreciprocity with one mirror axis.} }  (a) Linear transport response, $R_{\parallel}$ and $R_{\perp}$, and (b)  nonreciprocal transport response, \Vpo\ and \Vto, as a function of azimuth angle of current flow $\phi$ measured at $\nu = 1$. All measurements are performed at $T = 20$ mK, $B = 0$ and $I_{AC}=100$ nA. The mirror axis of the nonreciprocal response, marked by green solid lines in panel (b), is misaligned from those of the linear transport in panel (a). 
}
\end{figure*}

\subsection{The proximity effect with \WSe}

Fig.~\ref{figComparison} directly compares two tTLG samples with and without the proximity effect from a crystal of \WSe. Fermi surface reconstructions are observed in both samples near the same band fillings at $\nu=-2$, $+1$, $+2$ and $+3$ (Fig.~\ref{figComparison}a-d). In both samples, superconductivity emerges near half-filling of both electron- and hole-doped bands (Fig.~\ref{figComparison}a-b). Moreover, the $\nu-T$ maps from both samples reveal the same phase diagram for isospin polarized phases, which are identified based on the oscillation in the longitudinal resistance. This is a strong confirmation that the influence of the proximity effect is not essential in realizing the angular dependence of nonreciprocity.

The orbital ferromagnetic (OF) order, however, is only observed in the tTLG sample with the proximity effect with \WSe. First of all, the observation of the OF in tTLG/\WSe\ is consistent with previous observations in magic-angle twisted bilayer graphene, where anomalous Hall effect is enhanced by the proximity with \WSe\ ~\cite{Lin2022SOC}. The OF order requires simultaneously polarization in valley isospin and sublattice ~\cite{Serlin2019,Sharpe2019}. In the tTLG/\WSe\ sample, we propose that sublattice symmetry breaking is induced by the presence of \WSe, which accounts for the observation of the OF order in tTLG/\WSe\ heterostructure.

\begin{figure*}[!h]
\includegraphics[width=1\linewidth]{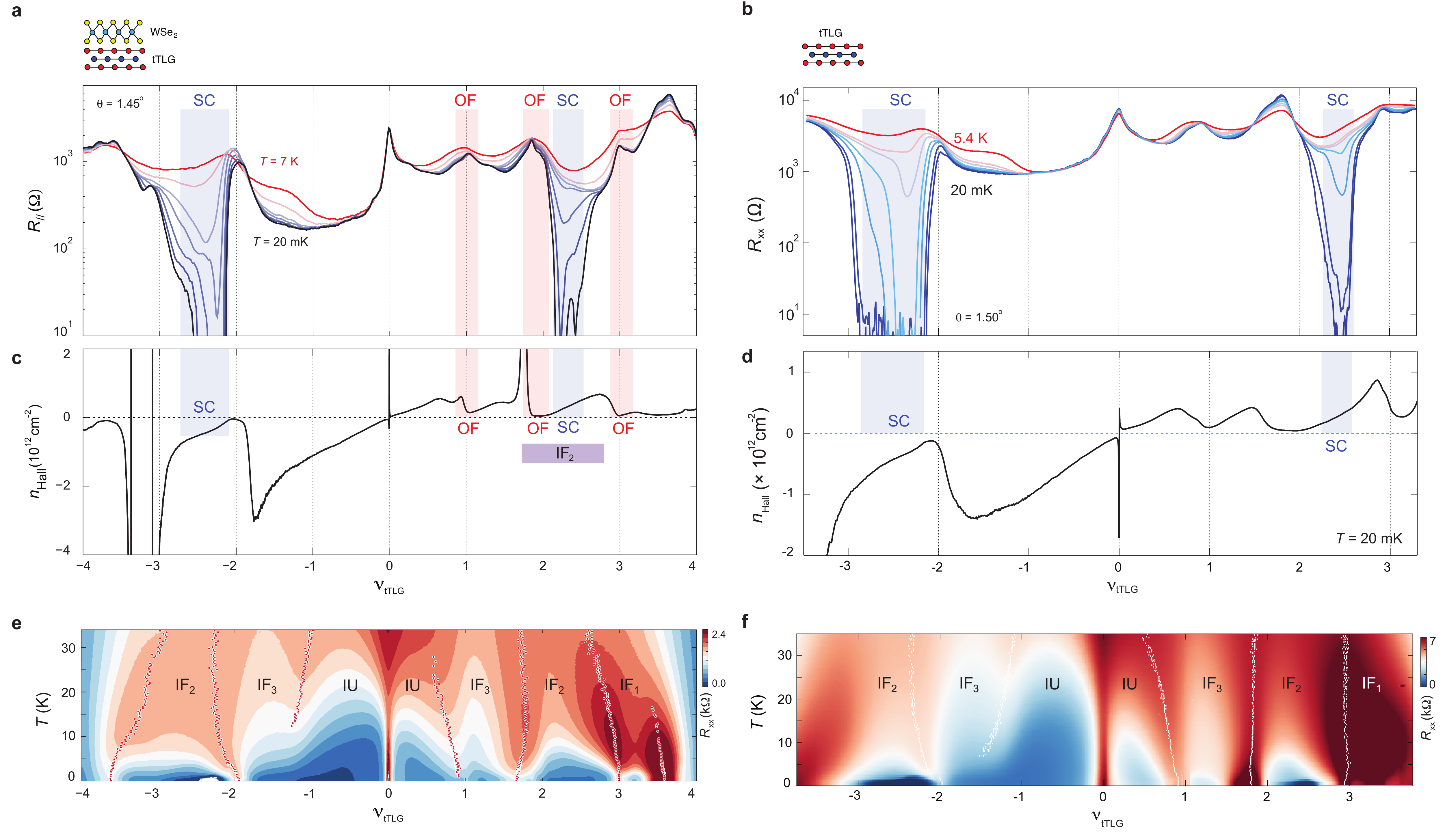}
\caption{\label{figComparison} {\bf{Comparison between tTLG samples with and without proximity effect.}}  (a-b) Longitudinal resistance and (c-d) Hall density as a function of the moir\'e filling dependence. Panels (a) and (c) are measured from the tTLG/\WSe\ heterostructure, which is used in the current work. Whereas panels (b) and (d) are measured from a tTLG sample without the prixmity effect, which is used in Ref.~\cite{Liu2022DtTLG}. Vertical dashed lines mark integer moir\'e band fillings, which coincide with resistance peaks in the longitudinal resistance, as well as resets in the Hall density. (e-f) Temperature-density map across the moir\'e band measured from tTLG samples (e) with and (f) without the proximity effect from \WSe. White circles mark the position of resistance maxima, which separates the $\nu-T$ map into regimes with distinct isospin polarization. The boundary between $IF_3$, isospin ferromagnet with $3$-fold degeneracy, and $IU$, isospin-unpolarized phases exhibit a  unique slope that is indicative of a Pomeranchuk-type transition ~\cite{Saito2021pomeranchuk}. 
 }
\end{figure*}

\subsection{Fermi surface contour and nonreciprocity}

\begin{figure}[h]
\includegraphics[width=0.7\linewidth]{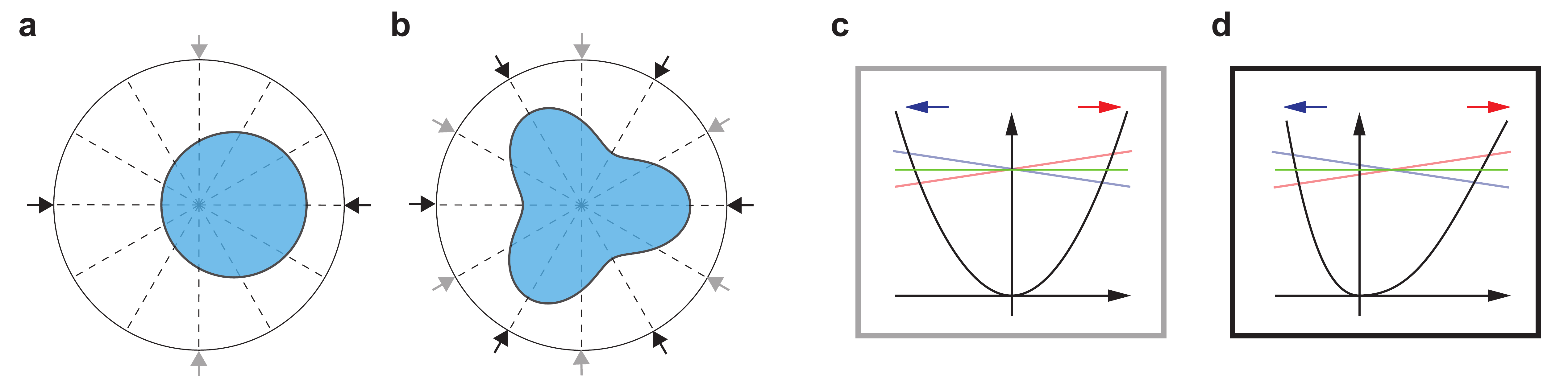}
\caption{\label{figband} {\bf{Fermi surface contour and band dispersion.}}  (a-b) Schematic diagram of distorted Fermi surface contours with (a) one and (b) three mirror axes. The blue shaded area denotes occupied electronic states within the Fermi surface. The edge of the energy band (zero momentum) is located at the center of the circle. The azimuth directions with symmetric and asymmetric band dispersion are marked with grey and black arrows, respectively.  (c) A symmetric and (d) asymmetric energy band structure relative to the bottom of the band. The Fermi surface tilts away from the equilibrium position (red and blue lines) in response of a large current bias along the direction of positive (red arrow) and negative (blue arrow) momentum. 
}
\end{figure}


In the presence of two-fold inplane rotational $C_2$ symmetry breaking, the Fermi surface is distorted, giving rise to a contour as shown in Fig.~\ref{fig1}b-c, as well as Fig.~\ref{figband}a-b. 

For a circular Fermi surface that preserves $C_2$, the band dispersion is symmetric with regard to $k=0$ for any azimuth direction, as shown in Fig.~\ref{figband}c. In this case, the same transport response is expected between forward and reversed current bias.
On the other hand, a distorted Fermi surface with one or three mirror axis features asymmetric band dispersion (Fig.~\ref{figband}d) along the direction of the mirror axis (marked by black solid arrows in Fig.~\ref{figband}a-b). When current flows along these directions, the transport response is expected to be highly nonreciprocal. 

Noticeably, along certain directions, even the distorted Fermi surface exhibits symmetric dispersion. These directions are marked by gray arrows in Fig.~\ref{figband}b-c. When current flows along these directions, nonreciprocity is expected to diminish to zero. This gives rise to the angular oscillation in nonreciprocity. 






The broken symmetries of a Fermi surface can be characterized by the azimuth directions where the energy band structure is symmetric, or asymmetric. As shown in Fig.~\ref{figband}b and c, the azimuth directions with symmetric and asymmetric band dispersion are marked with grey and black arrows, respectively. For instance, the one-fold symmetric Fermi surface exhibits an asymmetric band dispersion along azimuth direction of $\phi = 0$ and $180^{\circ}$, whereas the band structure is symmetric along $\phi = 90^{\circ}$ and $270^{\circ}$. This Fermi surface contour can be identified based on the angular dependence of transport nonreciprocity. Maximum positive and negative nonreciprocity is expected when current flows along  $\phi = 0$ and $180^{\circ}$, whereas the transport behavior should be reciprocal along $\phi = 90^{\circ}$ and $270^{\circ}$. On the other hand, the three-fold symmetric Fermi surface in Fig.~\ref{figband}c will give rise to a different angular dependence. There are three pairs of azimuth direction where prominent transport nonreciprocity is expected (black arrows). Transport nonreciprocity diminishes when current flows along the azimuth directions marked by grey arrows. Taken together, the ARNTM provides a direct tool to probe and characterize the shape of Fermi surface contour.

If charge carriers equally occupy three trigonal-warping-induced Fermi pockets in one of the valley, the symmetry of the resulting Fermi surface is equivalent to Fig.~\ref{figband}c. In this case, we expect to observe a three-fold symmetric angular dependence in the transport nonreciprocity. The exchange-driven instability in the momentum space induces charge carriers to condense into one of the Fermi pocket. The resulting Fermi surface contour shares the same broken symmetries as Fig.~\ref{figband}b, which can be identified based on a one-fold symmetric angular dependence in the transport nonreciprocity.

\subsection{Cascade of isospin transitions}

Fig.~\ref{figSIfan} shows a cascade of Fermi surface reconstructions at integer fillings of the moir\'e band. These reconstructions are evidenced by quantum oscillations emanating from integer fillings of the moir\'e bands (Fig.~\ref{figSIfan}a-c), as well as resets in $n_{Hall}$ (Fig.~\ref{figSIfan}d). The reconstruction gives rise energy gaps in the moir\'e band. Owing to the presence of a Dirac-like band in tTLG, gapped state in the moir\'e flatband exhibits a slope of $\pm2$ in the $\nu-B$ map, which are marked with solid green lines in Fig.~\ref{figSIfan}c.

\begin{figure*}[!t]
\includegraphics[width=0.7\linewidth]{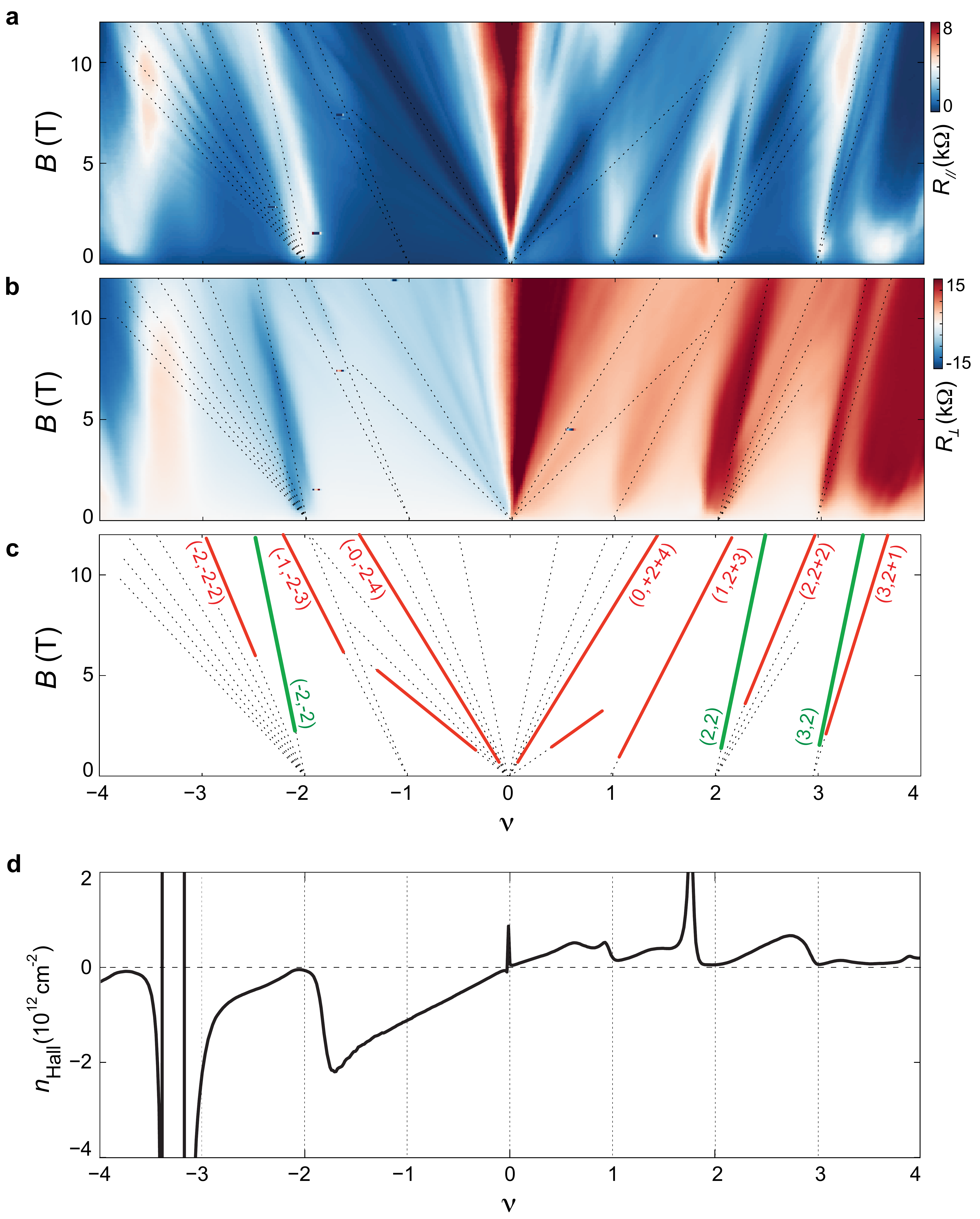}
\caption{\label{figSIfan} {\bf{Magneto-transport measurement across the moir\'e flatband.} }  
(a) Longitudinal resistance $R_\parallel$ and (b) Hall resistance $R_\perp$ across the moir\'e filling-magnetic field ($\nu-B$) map. Incompressible states are manifested as minima in $R_{\parallel}$, concomitant with quantized plateau in $R_{\perp}$. (c) The most prominent incompressible states are marked with black dashed line in the schematic $\nu-B$ map, where each trajectory is described by a pair of quantum numbers ($t, s$) from the Diophantine equation $\nu = t \phi/\phi_{0} + s$. Here $\nu$ is the moir\'e filling factor at the incompressible state, whereas $t$ and $s$ describe the slope and intercept of each trajectory ~\cite{Xie2021tblg,Spanton2018}. Near the CNP, the most prominent sequence of quantum oscillation exhibits a degeneracy of $4$. This indicates that the  underlying Fermi surface is isospin-unpolarized. However, the ARNTM at $B=0$ shows a non-zero nonlinear response, which points towards a partial valley- and momentum-polarization. 
(d) Hall density $n_{Hall}$ as a function of moir\'e filling measured at $B = 0.5$ T.
}
\end{figure*} 

The isospin degeneracy underlying the reconstructed Fermi surface can be determined based on the slope of the quantum oscillation in the fundamental sequence emanating from each integer moir\'e filling. The lowest order state of each fundamental sequence is marked by the solid red lines in Fig.~\ref{figSIfan}c. 
The trajectory of each incompressible state is described by a pair of quantum numbers ($t, s$) from the Diophantine equation $\nu = t \phi/\phi_{0} + s$, where $t$ corresponds to the slope and $s$ the intercept in the $\nu-B$ map. 
The isospin degeneracy of the Fermi surface equals the difference between slope of the Coulomb-driven energy gap in the moir\'e band, which is $\pm2$, and that of the lowest order quantum oscillation. For instance, the moir\'e energy gap at $\nu=2$ traces a slope of $t=2$, whereas the fundamental sequence of quantum oscillation has a slope of $t=2+2$.  This points to a degeneracy of $2$ for the reconstructed Fermi surface at half moir\'e filling. Similarly, the reconstructed Fermi surface at quarter and three-quarter fillings have $3$ and $1$ fold degeneracy. This hierarchy of isospin degeneracy is in excellent agreement with previous report in magic-angle twisted bilayer graphene ~\cite{Das2021symmetry,Saito2021symmetry}, suggesting that the underlying moir\'e band of twisted trilayer graphene that is directly comparable to the flatband physics studied in MATBG.

The Landau fan diagram in Fig.~\ref{figSIfan} provides an identification for the twist angle at $\theta = 1.44^{\circ}$. While the twist angle is slightly detuned from the magic angle of $\sim 1.55^{\circ}$, the emergence of Landau fan at each integer filling, along with Hall density reset, exhibit the same characteristic as magic-angle twisted trilayer graphene in previous observations ~\cite{Park2021tTLG,Hao2021tTLG,Liu2022DtTLG}. 

\section{Supplementary Data}

\begin{figure*}
\includegraphics[width=0.7\linewidth]{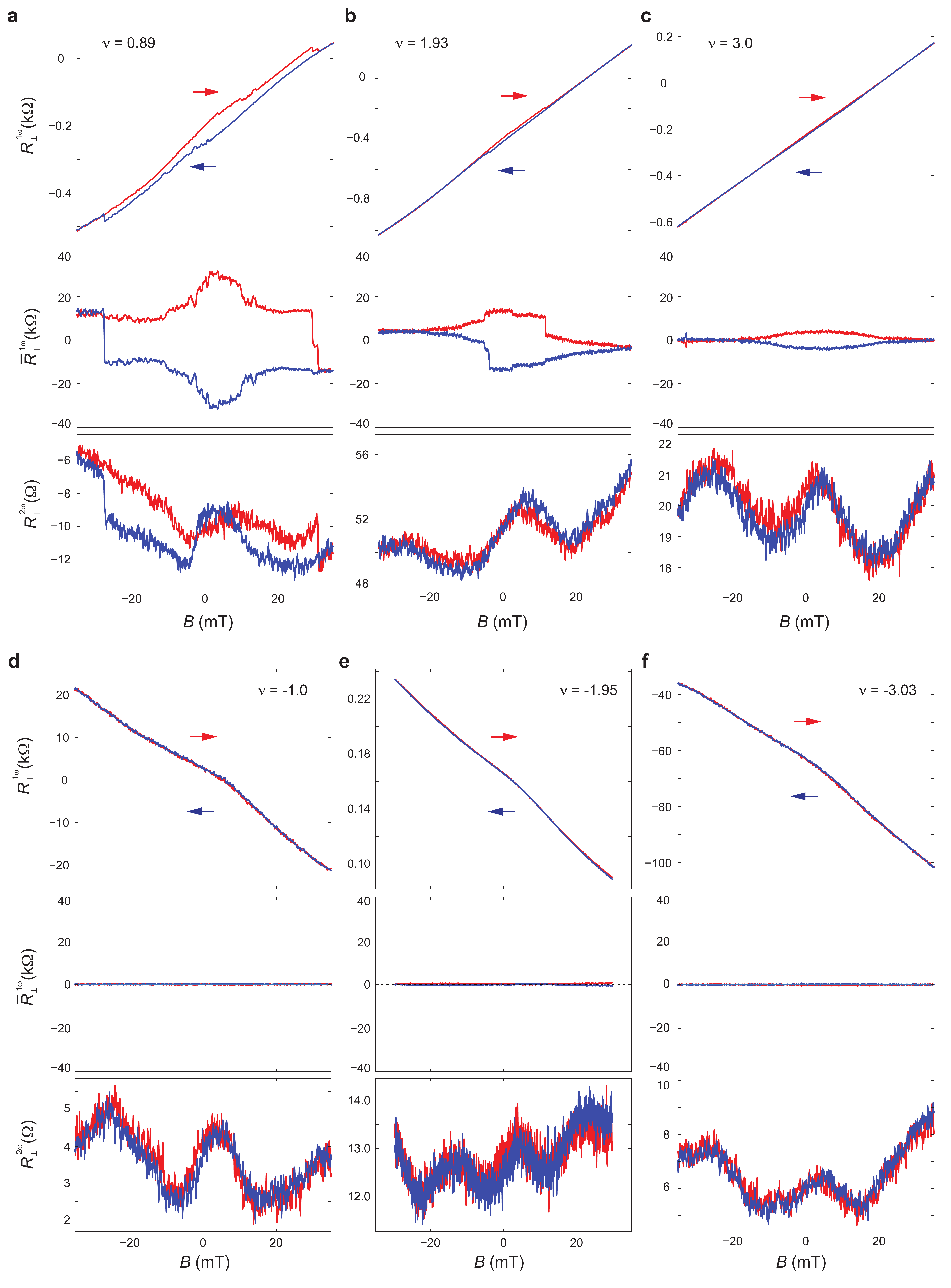}
\caption{\label{figLoop1} {\bf{$B$-induced hysteresis loops.} }  $B$-dependence of Hall resistance $R_\perp$ (top panel), renormalized Hall resistance $\bar{R}_{\perp}$ (middle panel), and \Vto\  measured at different moir\'e band fillings. $\bar{R}_{\perp}$ is obtained by subtracting the common $B$-dependent background from $R_\perp$. The small size of the magnetic hysteresis loop likely results from the the Dirac band of tTLG, which keeps the sample metallic while the flatband is gapped by the emergence of the OF order. All measurements performed at $T=40$ mK and $I_{AC}=100$ nA.
}
\end{figure*}

\begin{figure*}
\includegraphics[width=0.75\linewidth]{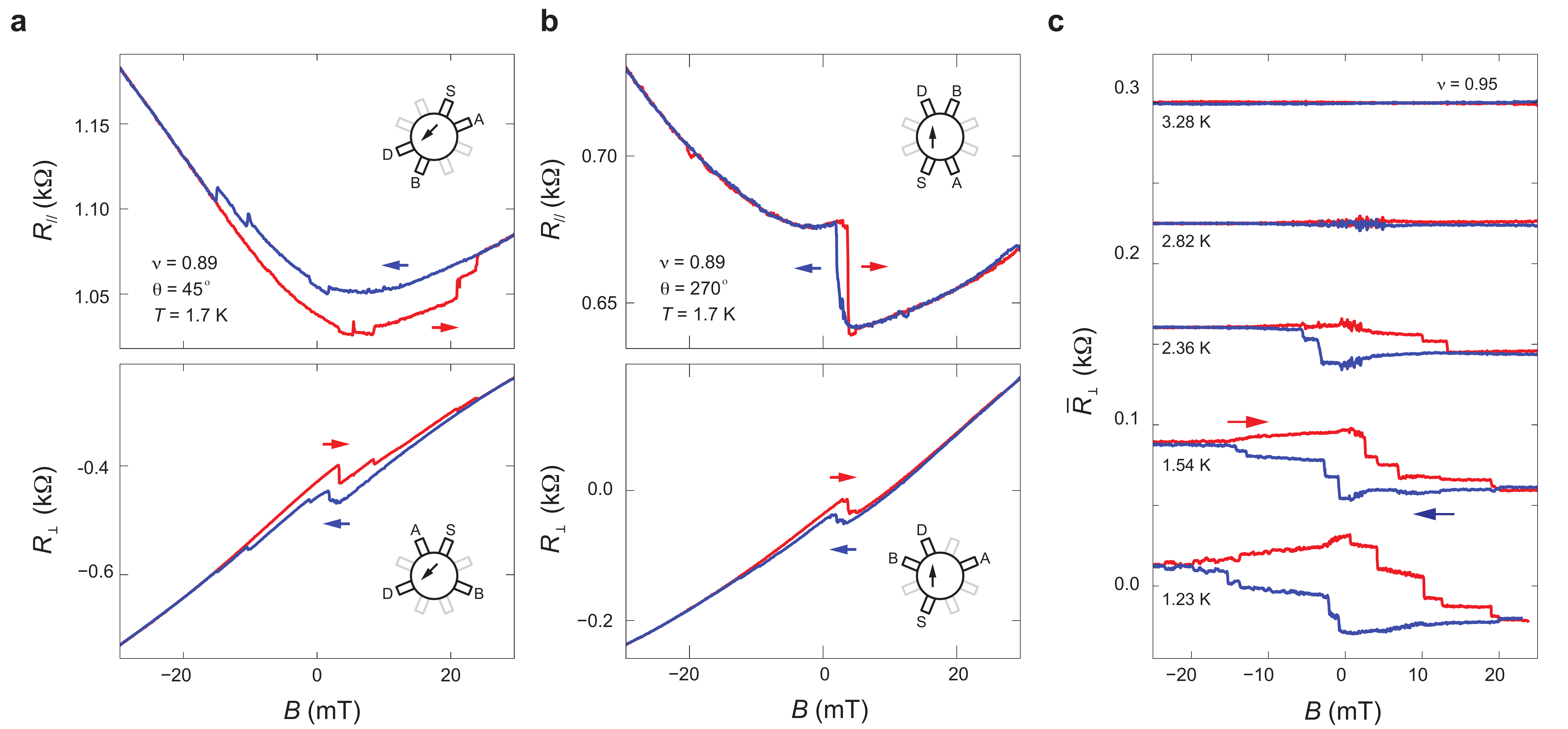}
\caption{\label{figBloopT} {\bf{$B$-induced hysteresis loops.} } (a-b) $B$-induced hysteresis loops in $R_{\parallel}$ (top panels) and $R_{\perp}$ (bottom panels) measured with DC current bias is applied along different azimuth directions (a) $\nu = 45^{\circ}$ and (b) $\nu = 270^{\circ}$. The hysteresis loop in $B$ shows dependence on the direction of current flow. While the angular dependence could simply arise from inhomogeneous domain distribution across the sample, the magnetic field is swept at the same rate. As such, the angle-depednent variation in the hysteresis behavior cannot be accounted for by an intrinsic hysteresis of the magnetic field. Moreover, the prominent jump in $R_{\parallel}$, as shown in the top panel of (b), is characteristic of a hysteretic transition in the underlying orbital ferromagnetic order. Combined with the current-indcued transitions in Fig.~\ref{fig4}b, we conclude that anomalous Hall effect is observed near integer filling $\nu=1$.  (c) Renormalized Hall resistance, $\bar{R}_\perp$, measured  at different temperature while $B$-field is swept back and forth. $\bar{R}_\perp$ is obtained by subtracting the $B$-dependent background from $R_\perp$. Measurements in (a-b) performed at $T=1.7$ K; all measurements performed with $I_{AC}=40$ nA. 
}
\end{figure*} 


\begin{figure*}
\includegraphics[width=0.8\linewidth]{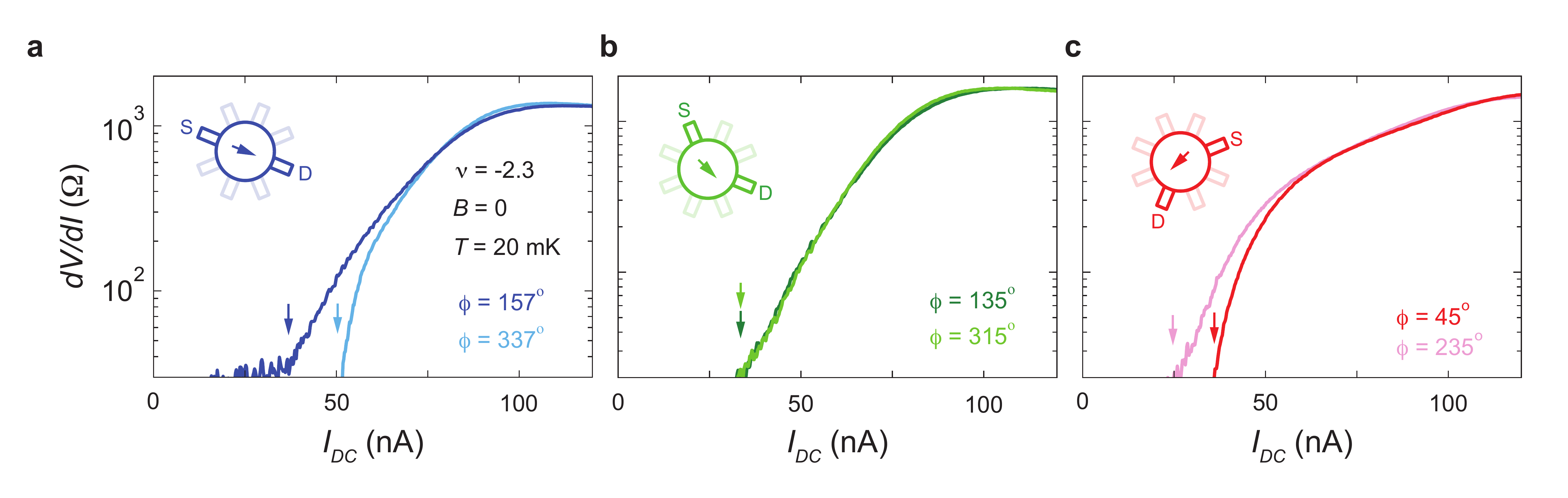}
\caption{\label{EtaIc} {\bf{Nonreciprocity in the critical supercurrent. }} 
Current-voltage (IV) characteristic of the superconducting phase measured at $\nu=-2.3$ with different current flow directions. The IV curve is measured with d.c. current flowing along (a) $\phi=157^\circ$, (b) $135^\circ$, and (c) $45^\circ$. Solid and dashed lines denote forward and reverse DC current biasing, respectively.  All measurements are performed at $\nu = -2.3$, $B = 0$ and $T = 20$ mK.  Vertical arrows mark the critical supercurrent, $I_c$, where the differential resistance $dV/dI$ becomes larger than the noise floor.  Black and grey arrows indicates the value of the critical supercurrent with forward and reverse current bias, respectively. The superconducting nonreciprocity is defined based on Eq.~4. Based on this definition, the value of $\eta_I$ is (a) negative, (b) zero, and (c) positive. Notably, all three IV curves are measured by slowly sweeping d.c. current from large reversed current bias to large forward current bias. The fact that different signs in $\eta_I$ are observed with the same sweep rate and direction provides a strong indication that the nonreciprocity in intrinsic to the superconducting phase. 
}
\end{figure*}




\begin{figure*}
\includegraphics[width=0.9\linewidth]{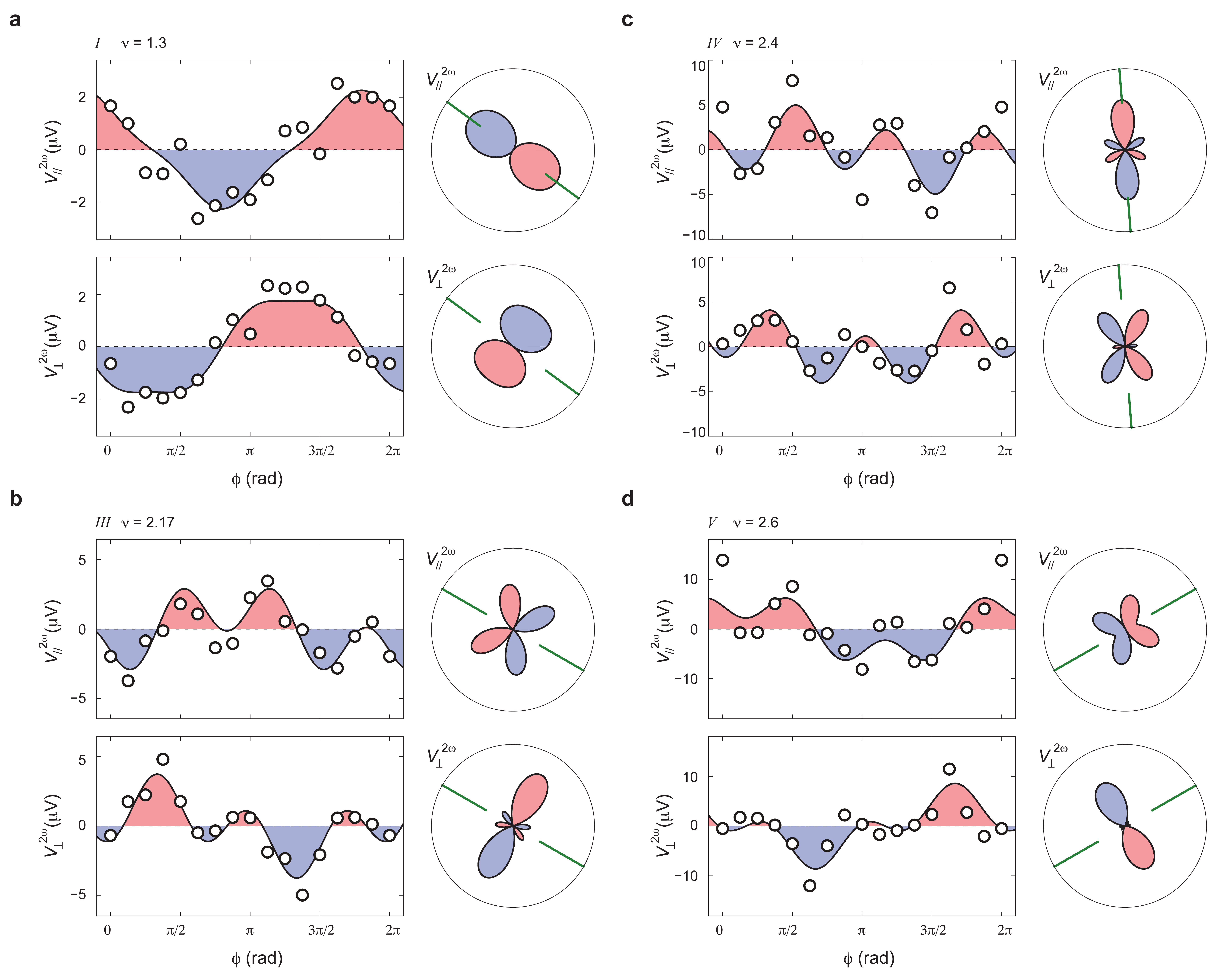}
\caption{\label{Cartesian3d} {\bf{Angular dependence of nonreciprocity shown in Fig.~\ref{fig3}d. }}
The angular dependence of\ \Vpo\ and \Vto\ shown in Fig.~\ref{fig3}d, plotted in Cartesian coordinates. All measurements performed at $T=20$ mK, $B=0$, and $I_{AC}=100$ nA.}
\end{figure*} 

\begin{figure*}
\includegraphics[width=0.9\linewidth]{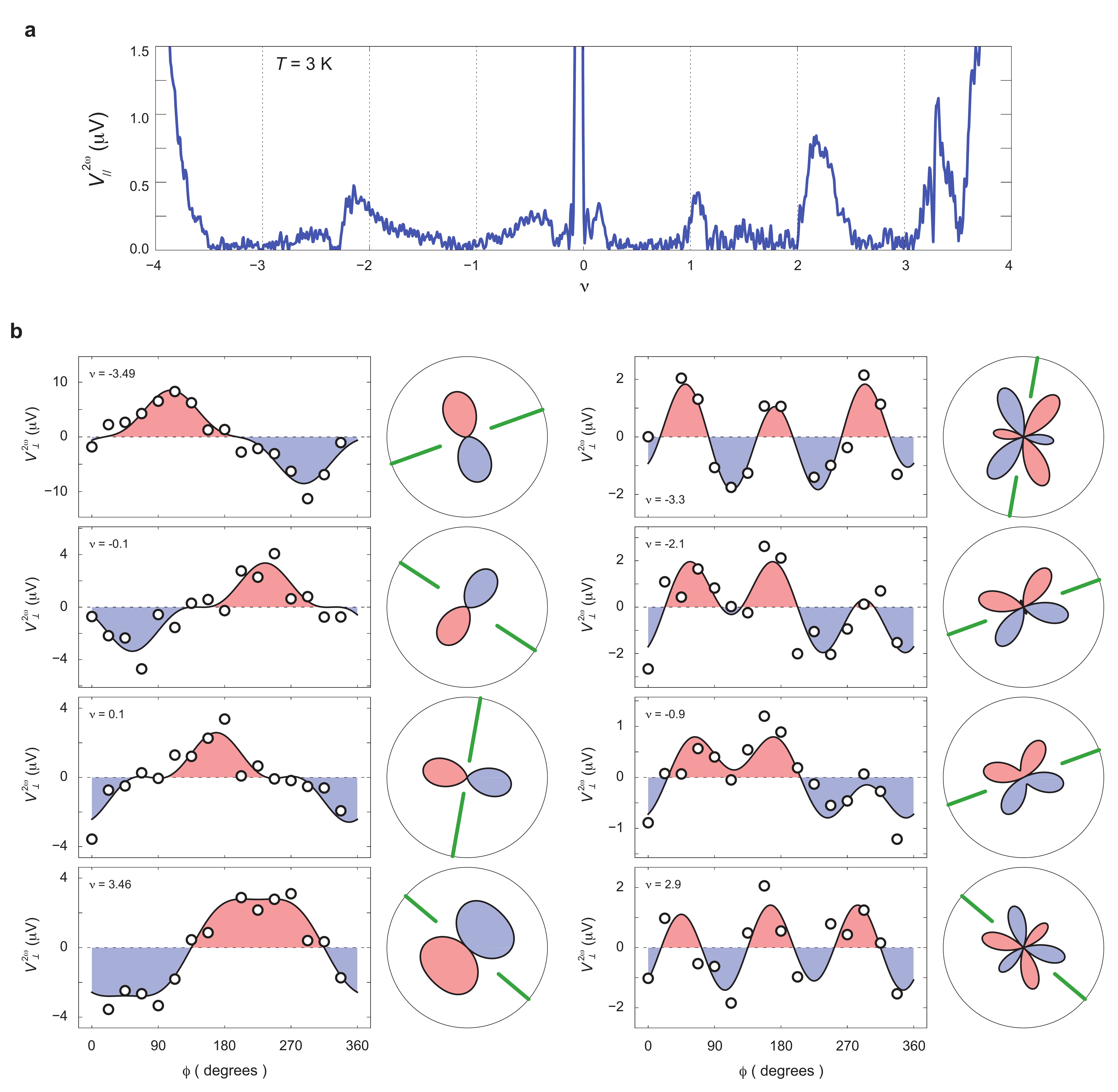}
\caption{\label{NonreciprocityMoire} {\bf{Angle-resolved transport nonreciprocity across the moir\'e band.}}
(a) \Vpo\ as a function of moir\'e doping measured at $T = 3$ K. At this temperature, nonreciprocal response is observed only near integer band fillings. This indicates that the strongest nonreciprocal response occurs near the cascade of Fermi surface reconstruction. Measurement is performed at $B=0$ with a current bias of $I_{AC}=100$nA. 
(b) The angular dependence of transport nonreciprocity measured at different moir\'e band fillings. Left panels show the Cartesian coordinates and right panels display polar-coordinate plots. All measurements are performed at $B=0$, $T=20$mK, and $I_{AC}=100$nA. The superconducting phase is suppressed by the large current bias.
}
\end{figure*}

\newpage
\clearpage

\begin{figure*}
\includegraphics[width=0.75\linewidth]{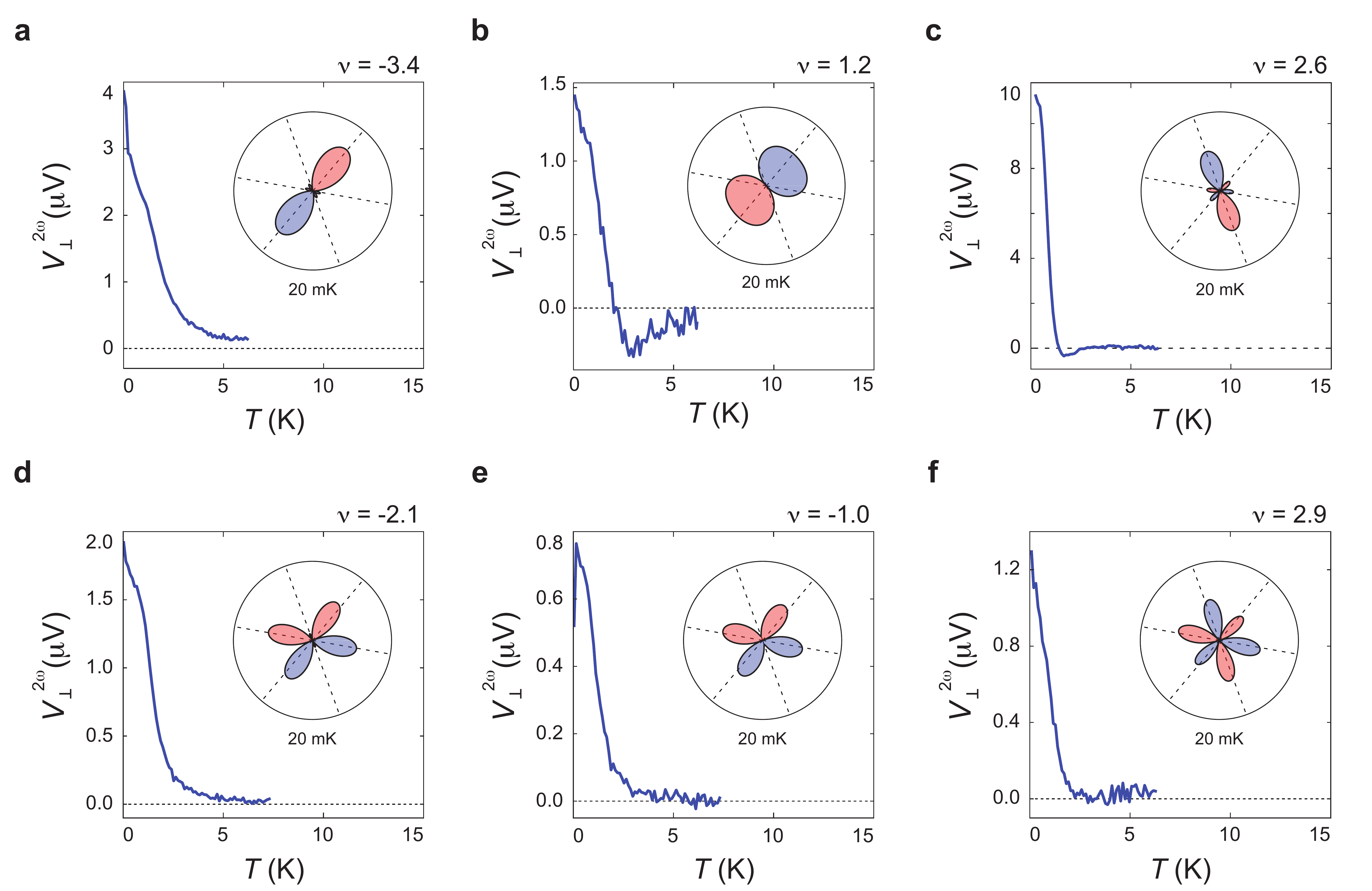}
\caption{\label{figSIT} {\bf{The temperature-onset of the loop current state.} }  
The temperature dependence of \Vto\ as a function of temperature measured at (a) $\nu=-3.4$, (b) $\nu=1.2$, (c) $\nu=2.6$,  (d) $\nu=-2.1$, (e) $\nu=-1.0$, and (f) $\nu=2.9$. In (a-c), the loop current phase exhibits one-fold symmetric \Vto.  In (d-f), the loop current phase exhibits predominantly three-fold symmetric \Vto. The onset of \Vto\ occurs at temperature below $5$ K, regardless of the one-fold or three-fold angular dependence. \Vto\  is the nonlinear response measured at the second-harmonic frequency with an AC current of $I_{AC} = 100$ nA. 
}
\end{figure*}

\begin{figure*}
\includegraphics[width=0.58\linewidth]{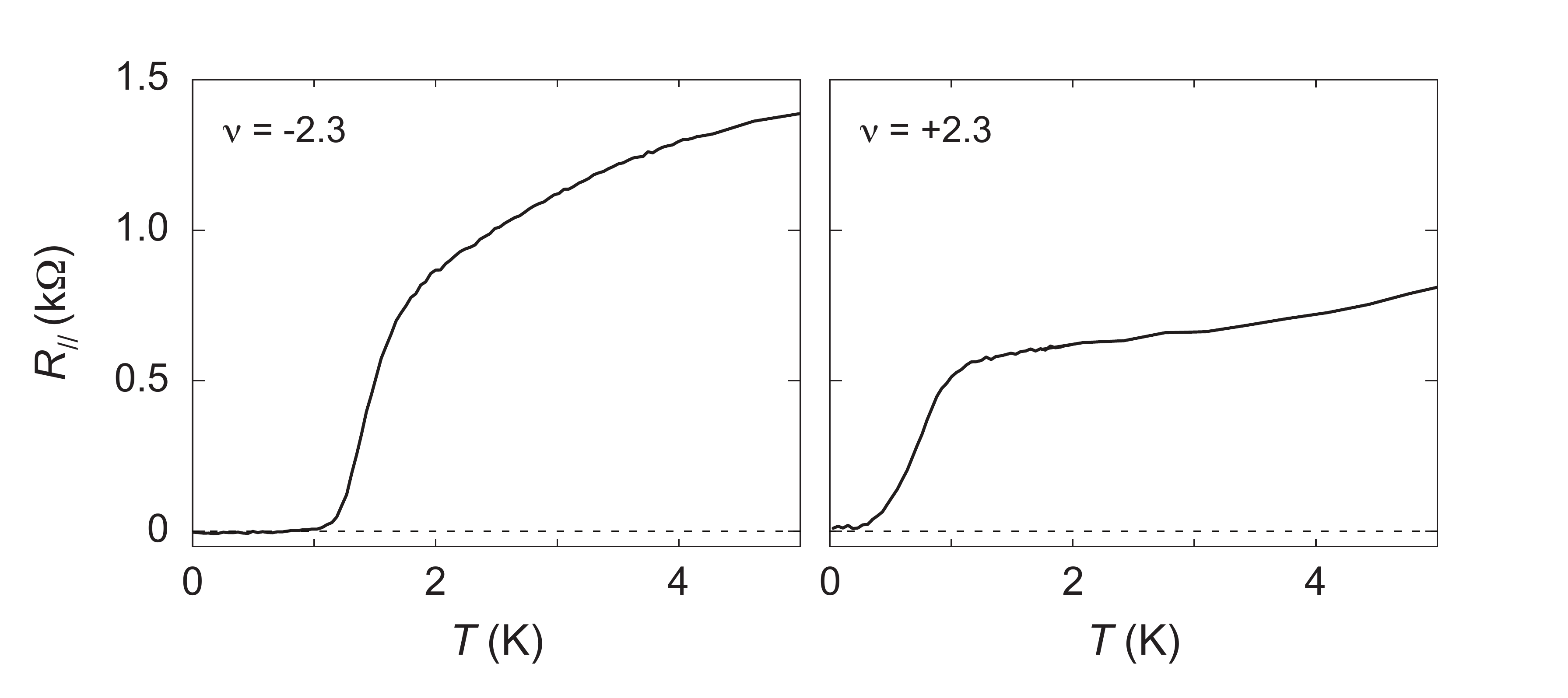}
\caption{\label{figSCRT} {\bf{Longitudinal resistance versus temperature for the superconducting phase.}} 
Longitudinal resistance as a function of temperature measured at $\nu=-2.3$ (left) and $\nu=+2.3$ (right). Measurement is performed at $B=0$ and $I =5 $nA. 
}
\end{figure*}

\newpage
\clearpage

\raggedbottom

\section{Materials and Method}

\subsection{Device Fabrication}

The doubly encapsulated \tlg\ is assembled using the “cut-and-stack” technique. All components of the structure are assembled from top to bottom using the same poly(bisphenol A carbonate) (PC)/polydimethylsiloxane (PDMS) stamp mounted on a glass slide. The sequence of stacking is: graphite as top gate electrode, 24 nm thick hBN as top dielectric, bilayer \WSe, \tlg, 24 nm thick hBN as bottom dielectric, bottom graphite as bottom gate electrode. The entire structure is deposited onto a doped Si/SiO$_2$ substrate, as shown in Fig.~\ref{figDevice}a. Electrical contacts to tTLG are made by CHF$_3$/O$_2$ etching and deposition of the Cr/Au (2/100 nm) metal edge contacts. The sample is shaped into an sunflower geometry with an inner radius of 1.9 $\mu$m for the circular part of the sample. In this geometry, the electrical contacts are separated by an azimuth angle of $45^\circ$, allowing an increment in the azimuth angle that is  $22.5^\circ$ (Fig.~\ref{figSetup}).

\begin{figure*}[h]
\includegraphics[width=0.85\linewidth]{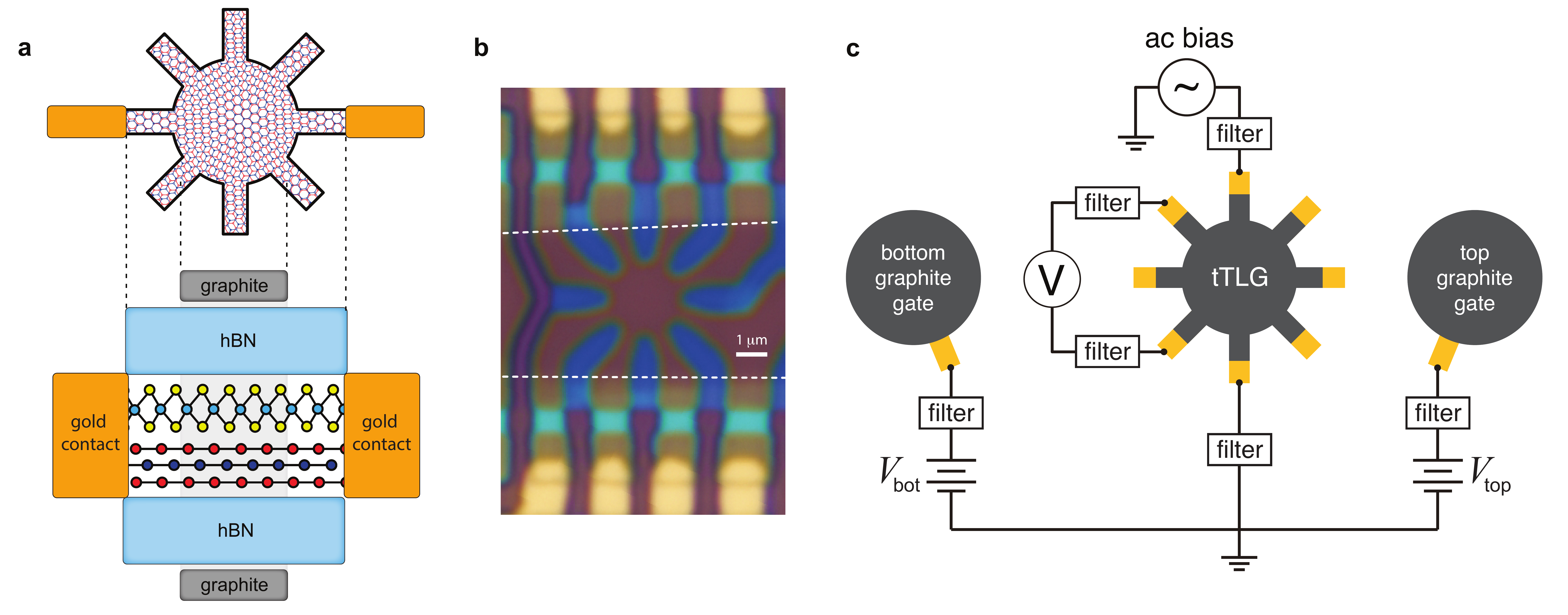}
\caption{\label{figDevice} {\bf{``sunflower'' geometry.} }   (a) Schematic of the ``sunflower'' geometry sample. (b) Optical image of the tTLG sample patterned into the ``sunflower'' geometry. 
Carrier density is varied by tuning the DC voltage bias applied to the bottom graphite gate, which only covers the circular part of the ``sunflower'' sample. Electrical contacts between graphene and metal are outside of the graphite gate. Therefore, eight ``patels''are fixed at a constant value of carrier density. As such, the dependence of linear and nonlinear transport responses on moir\'e band filling is decoupled from the potential influence of contact resistance.  Detailed characterization of transport response across such sample geometry is discussed in Ref.~\cite{Zhang2022sunflower}. (c) Schematic diagram of wiring used in the ARNTM. 
}
\end{figure*} 

\subsection{Transport measurement}

The carrier density in tTLG is tuned by applying a DC voltage bias to the bottom gate electrode. The electrical potential of the top gate electrode is held at zero. As a result, the tTLG sample experience a non-zero displacement field $D$ at large carrier density, which induces hybridization between the monolayer band and the moir\'e flatband. We note that the dependence of Hall density on moir\'e band filling is in excellent agreement with $D=0$ behavior from previous observations. This indicates that the influence of $D$ on the moir\'e flatband is not substantial. This is further confirmed by the Landau fan diagram in Fig.~\ref{figSIfan}, which is also consistent with the expected behavior at $D=0$. 

Transport measurement is performed in a BlueFors LD400 dilution refrigerator with a base temperature of 20 mK. Temperature is measured using a resistance thermometer located on the cold finger connecting the mixing chamber and the sample. An external multi-stage low-pass filter is installed on the mixing chamber of the dilution unit. The filter contains two filter banks, one with RC circuits and one with LC circuits. The radio frequency low-pass filter bank (RF) attenuates above 80 MHz, whereas the low frequency low-pass filter bank (RC) attenuates from 50 kHz. The filter is commercially available from QDevil. 

\pagebreak

\end{widetext}
\end{document}